%% file: d2dcoul.tex
\documentclass[12pt]{iopart}

\input newcom.tex

\begin{document}
\jl2

\title
%[Matrix elements in $d^2d'$ configuration]
[Coulomb and spin-orbit interaction matrix elements in $d^2d'$ configuration]
{Coulomb and spin-orbit interaction matrix elements in $\bi{d^2d'}$
configuration}
\author{Edwin Lo\footnote{e-mail: elo@pha.jhu.edu}}
\address{Henry A Rowland Department of Physics and Astronomy, The Johns
Hopkins University, Baltimore, Maryland 21218, USA}

\begin{abstract}
The $d^2d'$ configuration is analysed in group-theoretical terms.  Starting
from the table given by Condon and Odabasi for the configuration
$d^2d'$, we determine a set of convenient group-theoretical basis
states, and rewrite the Coulomb matrix elements in terms of this new basis.
Linear combinations from the different parts of the Coulomb operators are
formed such that they have simple group transformation properties in our
scheme.  The sequence of groups that we use is $U(20) \supset SO_T(3) \times
U(10) \supset SO_T(3) \times SO_S(3) \times U(5) \supset SO_T(3) \times
SO_S(3) \times SO(5) \supset SO_T(3) \times SO_S(3) \times SO_L(3)$, where
$T$ denotes the {\em isospin} of \v Simonis \etal, in which
electrons with the same angular momentum $l$ but different principle quantum
numbers $n$ are accommodated by introducing the eigenvalue $M_T$ of $T_0$.
Using the Wigner-Eckart theorem and selection rules on the higher symmetry
groups, the tables of the Coulomb and spin-orbit matrix elements for the
reconstituted operators (with simple group transformation properties)
are much simplified in terms of these basis states.
\end{abstract}

%\pacs{31.15.+q}
%\submitted
%\maketitle
%\newpage

\section{Introduction}

The configuration $d^2d'$ has been observed in the excited spectra of some
rather common atoms and ions.  For instance, from the data of Sugar and
Corliss (1985), Sc {\scriptsize I}, Ti {\scriptsize II}, V {\scriptsize III}
and Cr {\scriptsize IV} all have excited levels with the configuration
$3d^24d$ observed.  In the case of vanadium and chromium, almost all the 27
possible spectroscopic terms of the configuration $3d^24d$ were found and
their energies measured; for vanadium, a few $3d^25d$ levels were seen as well.
In addition to the $d^2d'$ systems, the present work also applies to commonly
found systems of the type $d^8d'$.  For example, $3d^84d$ has been observed
in Co {\scriptsize I} and Ni {\scriptsize II} (Sugar and Corliss, 1985).  For
nickel, all 27 possible spectroscopic terms of the configuration $3d^84d$ have
been found, as well as some levels of $3d^85d, 3d^86d, 3d^87d$ and $3d^88d$.

Condon and Odabasi (1980) calculated the Coulomb matrix elements of $d^2d'$
using classical methods.  Their basis couples the two equivalent
electrons ($d^2$) to an intermediate state $^{[S']}L'$, to which the $d'$
electron is coupled to give the final state $^{[S]}L$.  The basis is denoted
by $|(^{[S']}L') ^{[S]}L \rangle$, where $[S] \ = \ 2S+1$, the multiplicity.
This basis is quite good, as the $d'$ electron usually interacts rather
weakly with the other $d$ electrons. As one can check from the data for
V {\scriptsize III} and Cr {\scriptsize IV} (Sugar and Corliss, 1985), there
is not a substantial mixing of the states in this basis except for
$^2P,{}^2D$ and $^2F$.

Instead of this physical basis, we choose to use a different one that has
simple transformation properties under higher symmetry groups.  There are
advantages in doing so, particularly when they are taken as intermediate
states in a perturbation expansion (see, for example, Hansen \etal 1997). In
that case, the sum over intermediate states that appear render the choice of
basis irrelevant to the physical properties being investigated. And certainly,
if we introduce higher symmetry groups into the analysis, even though they
might not give ``good'' quantum numbers, in the sense that they do not
commute with the full Hamiltonian, they give simple results as far as the
computation of matrix elements is concerned.  The selection rules, and
especially the Wigner-Eckart theorem, when applied to higher groups,
strikingly reduce the amount of calculation and expose new mathematical
structure.

Other than matrix elements among the states of the $d^2d'$ configuration
itself, our interest in configuration interaction (CI) means that we will
also need the matrix elements that connect $d^2d'$ to $d^3$.  The physical
basis that is commonly used for $d^3$ coincides with the group-theoretical
basis that utilizes the higher unitary and symplectic groups $U(10)$ and
$Sp(10)$.  For this reason, it is best to use the same group-theoretical
basis for the configuration $d^2d'$ as that needed to study the configuration
interaction between $d^3$ and $d^2d'$.  Once again, the Wigner-Eckart theorem
and selection rules for the higher symmetry groups are invaluable.

The isospin formalism (\v Simonis \etal 1984, Kaniauskas \etal 1987) that we
will introduce into our work also provides a very useful setting for the
study of CI.  In the isospin formalism, the states in
$d^3$ and $d^2d'$ that carry the same group labels belong to the same isospin
multiplet, but with different isospin $z$-component $M_T$.  We can therefore
easily obtain the CI matrix elements from the ones for
$d^2d'$ if everything, the states and the operators, have simple
transformation properties under the symmetry groups.  These considerations
give us strong reasons to abandon the physical basis for $d^2d'$ and turn to
the group-theoretical one instead.

\section{The basis states\label{sec:states}}
The classification used in Condon and Odabasi (1980), in group-theoretical
language, is
\bea
SO_S^d(3) \times SO_L^d(3) \times SO_S^{d'}(3) \times SO_L^{d'}(3) \supset
SO_S(3) \times SO_L(3).
\eea
As explained earlier, this provides a set of good physical states, but there
are also good reasons to use an alternative basis which carries a set of
simple group labels, that is, irreducible representations (irreps).  The
groups that we have chosen are
\bea
U(10) \supset SO_S(3) \times U(5) \supset SO_S(3) \times SO(5) \supset
SO_S(3) \times SO_L(3).
\label{eq:subgp}\eea
The groups $U(5)$ and $SO(5)$ are based on the five orbital states of a $d$
(or $d'$) electron; the group $U(10)$ utilizes the spin of the electron as
well.  Another more common alternative used in configurations of equivalent
electrons is the sequence
\bea
U(10) \supset Sp(10) \supset SO_S(3) \times SO(5) \supset
SO_S(3) \times SO_L(3)
\label{eq:subgpsp}\eea
where the symplectic group $Sp(10)$ is closely related to the seniority
scheme widely used in atomic spectroscopy.  It is, however, felt that the
idea of seniority might not as useful in configurations with inequivalent
electrons.  At least not in the way we used it in configurations of
equivalent electrons.

\input states.tex

To simplify the notation, we use square brackets for unitary group labels,
with a subscript to denote the order of the group.  The zeros in the labels
will either be omitted or abbreviated as a dash (--).  Also, we write the
negative values with an overline to save space.  For example, the irreps
$[210^8],[1^20^6-1^2]$ of $U(10)$ and $[11100]$ of $U(5)$ will be denoted as
\urs{21}{10}, \uro{11}{10} and \urs{111}5 respectively. The exponent form is
retained if the same number (other than zero) occurs more than three times.
The advantages of using such a notation will become clear when we discuss the
generic branching rules of the type $U(2n) \rightarrow SU(2) \times U(n) \cong
SO(3) \times U(n)$ in table~\ref{tb:ugpbr}; as well as to tabulate Kronecker
products for unitary groups of any order $n$ (see table~\ref{tb:ucross}).

The mapping of states from the classical basis to the new basis was worked
out by Judd (1997) and the author, and is given in table~\ref{tb:states}.
The reader should for the moment disregard the first superscript labeling
each state.  That is the {\em isospin} label which will be introduced later
in section~\ref{sec:iso}.  If the alternative scheme involving
the symplectic group $Sp(10)$ is used instead, almost all of the states in
the first column of table~\ref{tb:states} corresponds to a single state in
that scheme.  The only mixing comes from a pair of $^2D$ states that share
the same labels in $U(10)$ and $SO_S(3) \times SO(5)$.  They are given on the
last two lines in table~\ref{tb:states}.

\section{The Coulomb interaction matrix elements}
The matrix element table given in Condon and Odabasi (1980)\footnote{A
negative sign for the matrix element
$\langle (^3F) ^2H|g_0|( ^1G) ^2H\rangle$
is missing in their table; its correct value should be -105.} can be
transformed to the new basis without difficulty. At the same time, we also
followed Judd (1998) and Racah (1954) and take specific linear combinations
of the two-electron Coulomb operators so that they transform irreducibly
under the action of the group $SO(5)$.

The full two-body Coulomb interaction contains three distinct parts.  One
involves only the $d$-electrons (or only the $d'$-electrons, if we had more
than one $d'$-electron.)  The other two are the direct and exchange parts
that involve one $d$ and one $d'$ electron.

We can write the perturbing Hamiltonian as
\bea
H_1 & = & \sum_k f_k(d,d) F^k(d,d) + f_k(d,d') F^k(d,d') + g_k(d,d')
G^k(d,d') \nonumber \\
&=& \sum_i e_i E^i+ \fb i \Fb i + \gb i \Gb i
\label{eq:h1} \eea
where the sums run over $k=0,2,4$ and $i=0,1,2$.  The $F^k$ and $G^k$ are the
usual direct and exchange radial integrals, and the matrix elements of
$f_k(d,d')$ and $g_k(d,d')$ in $d^2d'$ in the classical basis are tabulated
by Condon and Odabasi (1980).  The operators $e_0, e_1$ and $e_2$ on the
second line, constructed from the operators $f_k(d,d)$ by Judd (1998) and
Racah (1954) independently, transform irreducibly as (00), (00) and (22)
of $SO(5)$ respectively.  They are given by
\bea
e_0 & = f_0(d,d) \nonumber \\
e_1 & = \frac75 [f_0(d,d)+5f_2(d,d)+9f_4(d,d)] \nonumber \\
e_2 & = 63[f_2(d,d)-f_4(d,d)]
\label{eq:ftoe}\eea
and the corresponding $E^i$ are
\bea
E^0 & = F_0(d,d) -\frac72F_2(d,d) -\frac{63}2F_4(d,d) \nonumber \\
E^1 & = \frac52F_2(d,d) + \frac{45}2F_4(d,d) \nonumber \\
E^2 & = \frac12F_2(d,d) - \frac52F_4(d,d)
\label{eq:FtoE}\eea
where $F_0 = F^0, F_2 = F^2/49$ and $F_4 = F^4/441$ as defined by Condon
and \mbox{Shortley (1953)}.  Replacing $f_k(d,d)$ by $f_k(d,d')$
and $g_k(d,d')$ in equation~\eref{eq:ftoe}, we arrive at the operators
\fb i and \gb i respectively; and they transform as (00) and (22) in
$SO(5)$ just like the operators $e_i$.  Similar substitution in
equation~\eref{eq:FtoE} gives us the
relation between \Fb i and the direct radial integrals $F^k(d,d')$; and
likewise, the relation between \Gb i and $G^k(d,d')$.

In terms of the creation and annihilation operators, or rather in terms of
\mbox{$\bi v^{(k)}_{dd}=\sqrt2\bi w^{(0k)}_{dd}=-\sqrt2
(\bi{d^\dag d})^{(0k)}$}, the operators $e_i$ are given by
\numparts\bea
e_0 &=\frac52 : \bi v^{(0)}_{dd} \cdot\bi v^{(0)}_{dd} : \label{eq:e0} \\
e_1 &=\frac72 : \bi v^{(0)}_{dd}\cdot\bi v^{(0)}_{dd} : \ + \ : \bi
v^{(2)}_{dd}\cdot\bi v^{(2)}_{dd} : \ + \ : \bi v^{(4)}_{dd}\cdot\bi
v^{(4)}_{dd} : \label{eq:e1}\\
e_2 &=9 : \bi v^{(2)}_{dd}\cdot\bi v^{(2)}_{dd} : \ - \; 5 \ : \bi
v^{(4)}_{dd} \cdot\bi v^{(4)}_{dd} : \label{eq:e2}
\eea\endnumparts
where colons denote normal ordering\footnote{The colon notation is widely
used in Quantum Field theory; see Weinberg (1995) for example.  Lindgren and
Morrison (1981) also used normal ordering in atomic theory, and denote it by
curly brackets.  We prefer colons to avoid confusion with anti-commutation
brackets and brackets in general.}.
For instance, with Einstein summation
convention over repeated Greek indices ($\mu, \nu$ \mbox{etc.}) used
throughout the paper, the operator $e_0$ from the above can be written as
\numparts\bea\fl
e_0=\frac52 : \bi v^{(0)}_{dd}\cdot\bi v^{(0)}_{dd} : \ = \frac12
: (d^\dag_\mu d_\mu) \cdot (d^\dag_\nu d_\nu): \ = -\frac12(d^\dag_\mu
d^\dag_\nu d_\mu d_\nu) = \frac12(d^\dag_\mu d^\dag_\nu d_\nu d_\mu) \ .
\label{eq:e0create}\eea

The direct and exchange operators, \fb i and \gb i, also admit similar
expressions in terms of the operators $\bi v^{(k)}_{dd'} (=
-\sqrt2(\bi{d^\dag d'})^{(0k)})$, $\bi v^{(k)}_{d'd}$ and $\bi
v^{(k)}_{d'd'}$.  For example, \fb0 and \gb0 are given as
\bea
\fb0 &=5 : \bi v^{(0)}_{dd}\cdot\bi v^{(0)}_{d'd'} : \ &=
(d^\dag_\mu d'^\dag_\nu d'_\nu d_\mu) \label{eq:f0create}\\
\gb0 &=5 : \bi v^{(0)}_{dd'}\cdot\bi v^{(0)}_{d'd} : \ &=
(d^\dag_\mu d'^\dag_\nu d_\nu d'_\mu) \label{eq:g0create}\ .
\eea\endnumparts
The extra factor $\frac12$ in equations~\eref{eq:e0} and~\eref{eq:e0create}
is to allow for the pairwise interaction between the identical $d$-electrons
being counted twice.  The other direct and exchange operators \fb i and \gb i
are similarly given as in equations~\eref{eq:e1} and~\eref{eq:e2}, with an
extra factor of two.  As a note, normal ordering removes the
self-interaction terms in $e_i$; and it take cares of the ambiguity that
arises in the case of the exchange operators \gb i, in which $\bi
v^{(k)}_{dd'}\cdot\bi v^{(k)}_{d'd}$ and $\bi v^{(k)}_{d'd}\cdot\bi
v^{(k)}_{dd'}$ are both legitimate but distinct forms, if not normal ordered.

In terms of the new basis states given in table~\ref{tb:states}, the matrix
elements of the operators $e_i$, \fb i and \gb i can be easily obtained from
the table by Condon and Odabasi (1980).  We simply transform their matrix
elements\footnote{Notice that the matrix elements of $f^k(d,d)$ in the table
of Condon and Odabasi (1980) are simply taken from the configuration $d^2$.}
of $f_k(d,d), f_k(d,d')$ and $g_k(d,d')$ into the new basis, then take the
corresponding linear combinations as given in equation~\eref{eq:ftoe}.  In
this way, we arrive at table~\ref{tb:matrix}.

\input tbmatrix.tex

\subsection{The irreps of the Coulomb operators}
As can be seen from table~\ref{tb:states}, all the states have definite group
labels with respect to the symmetry groups in~\eref{eq:subgp}. We have now
to determine the transformation properties of the nine operators $e_i$, \fb i
and \gb i.  Of course, we know that they are scalars in the spin and orbital
spaces.  As mentioned in the previous section, these nine operators thus
constructed belong to either (00) or (22) of $SO(5)$.  We will now
find the $U(5)$ and $U(10)$ labels of these operators.

The two-body operators contain two creation and two annihilation operators.
As the creation operators ($\bi d'^\dag$ or $\bi d^\dag$) transform as
\urs15, while annihilation operators \ursb15, the product
\[ \fl \urs15 \times \urs15 \times \ursb15 \times \ursb15 = 2\urs05 +
4\uro15 + \uroh2{11}5 + \uroh{11}25 + \uro{11}5 + \uro25 \]
must contain the appropriate labels for the two-body operators.  We only need
to pick out those $U(5)$ irreps from the Kronecker product above that contain
either (00) or (22) of $SO(5)$.  The branching rules $U(5) \rightarrow
SO(5)$ are given by Wybourne (1970)\footnote{The table by Wybourne (1970)
does not include $U(5)$ irreps with negative values.  But if we look at the
reduction $U(5) \rightarrow SU(5) \rightarrow SO(5)$ instead, the
fact that $[a_1,a_2,\cdots]$ and $[a_1+c,a_2+c,\cdots]$ of $U(5)$ reduce to
the same $SU(5)$ multiplet tells us they must contain the same $SO(5)$
components.  So, we can easily remove the negative numbers in each irrep of
$U(5)$ without changing its $SO(5)$ content.},
for instance.  We are then left with \urs05, \uro{11}5
and \uro25 where the first irrep contain (00) of $SO(5)$, the second one
contain (22) and the last one both.

On the $U(10)$ level, we use the branching rules given in table~\ref{tb:ugpbr}
and pick out the irreps in the Kronecker product
$\urs1{10} \times \urs1{10} \times \ursb1{10} \times \ursb1{10}$ that
contain either \surs105, \suro1{11}5 or \suro125 of $SO_S(3) \times U(5)$.
(The Coulomb operators are spin scalars.)  The eligible ones are
\urs0{10}, \uro{11}{10} and \uro2{10}.

The group labels for the nine operators $e_i$, \fb i and \gb i can thus be
written down rapidly as given in table~\ref{tb:oplabel}.  (Once again, the
reader can ignore the isospin superscript for the time being.)  The only ones
that need explanation are the three $e_i$ operators as well as \fb0 and \gb0.
The three equivalent electron operators $e_i$ do not contain the irrep
\uro2{10} of $U(10)$ because the two creation operators (and likewise, the
annihilation operators) must form an antisymmetric product in $U(10)$. So no
\uro2{10} appears.  In fact, they were examined by Judd and Leavitt (1986),
and it is found that $e_0$ belongs to \urs0{10} of $U(10)$, $e_2$ belongs
to \uro{11}{10}, and $e_1$ is a mixture of \urs0{10} and \uro{11}{10}.

For the operators $e_0$, \fb0 and \gb0, they can be shown to be $U(10)$
scalars rather easily.  From the expressions~\eref{eq:e0create}
through~\eref{eq:g0create}, it is a simple exercise to show
that they commute with all the 100 generators
$d^\dag_\mu d_\nu + d'^\dag_\mu d'_\nu$ of the group $U(10)$.
(See section~\ref{sec:iso} for further discussion of the group generators.)
Hence, they are all $U(10)$ scalars.

\input ubranch.tex

\input oplabel.tex

\subsection{Selection rules and the Wigner-Eckart Theorem\label{sec:selection}}
With the group labels for the operators worked out, we are now in a
position to appreciate the use of the new group-theoretical basis.  We can
look at the situation on different group levels.  First of all, the six
$SO(5)$ scalars operators $e_0, e_1$, \fb0, \fb1, \gb0 and \gb1 cannot
connect states with different $SO(5)$ labels.  And for the other three
operators $e_2$, \fb2 and \gb2 which belong to (22), the Kronecker products
for $SO(5)$ account for all the vanishing matrix elements one can find in
table~\ref{tb:matrix} on those three columns.

When we apply the Wigner-Eckart theorem on the group $U(10)$,
we expect a lot of simple proportionality relations among the different
operators.  Between the operators \fb i and \gb i, we found
\numparts\bea
\langle\urs{111}{10}\beta |\fb i|\urs{111}{10}\gamma\rangle & = &
\langle\urs{111}{10}\beta |\gb i|\urs{111}{10}\gamma\rangle \label{eq:WEfg1}\\
\langle\urs{111}{10}\beta |\fb i|\urs{21}{10}\gamma\rangle & = &
\langle\urs{111}{10}\beta |\gb i|\urs{21}{10}\gamma\rangle
\label{eq:WEfg2}\eea\endnumparts
for $i=0,1$ and 2; and $\beta, \gamma$ denotes the rest of the quantum numbers
needed to specify the state.  As for the relations with the operators $e_i$,
we have
\numparts\bea
\langle\urs{111}{10}\beta |\fb i|\urs{111}{10}\gamma\rangle & = & 2 \
\langle\urs{111}{10}\beta |e_i|\urs{111}{10}\gamma\rangle \label{eq:WEefg1}\\
\langle\urs{111}{10}\beta |\fb i|\urs{21}{10}\gamma\rangle & = &
-\langle\urs{111}{10}\beta |e_i|\urs{21}{10}\gamma\rangle \label{eq:WEefg2}\\
\langle\urs{21}{10}\beta |\fb i+\gb i|\urs{21}{10}\gamma\rangle & = &
\langle\urs{21}{10}\beta |e_i|\urs{21}{10}\gamma\rangle \ .
\label{eq:WEefg3}\eea\endnumparts

Most of the proportionality relations above are the result of the
Wigner-Eckart theorem at work on the group $U(10)$.  To demonstrate, let us
use a few examples.  For $i=2$ in equation~\eref{eq:WEfg1}, since
$\ursb{111}{10} \times \urs{111}{10}$ contains \uro{11}{10} once and no
\uro2{10} appears (see table~\ref{tb:ucross}), only the \uro{11}{10} part
(but not the \uro2{10} part) of the operators \fb2 and \gb2 contributes to
their matrix elements.  Hence, by the Wigner-Eckart theorem, they should be
proportional to each other, and the proportionality constant turns out to be
unity. Analogous argument applies to explain the proportionality relations
in~\eref{eq:WEfg2},~\eref{eq:WEefg1} and~\eref{eq:WEefg2} for $i=0,2$.
The $i=1$ operators contain the irreps \urs0{10} and \uro{11}{10}, which both
contribute to the matrix elements in general.  So, the Wigner-Eckart theorem
fails to provide an explanation for these $i=1$ cases in~\eref{eq:WEfg1}
and~\eref{eq:WEefg1}; although it works well with~\eref{eq:WEfg2}
and~\eref{eq:WEefg2} where the scalar part \urs0{10}
does not contribute to the matrix elements.  But in any case, it cannot
explain why the proportionality constants are the same for $i=0,1$ and 2.

The last relation~\eref{eq:WEefg3} is a little bit different.  Since
$\ursb{21}{10} \times \urs{21}{10}$ contain both \uro{11}{10} and \uro2{10},
we do not have a simple relation as we had before.  For $i=2$, if we write
equation~\eref{eq:WEefg3} as $\langle\gb2\rangle = \langle e_2 \rangle -
\langle \fb 2\rangle$, we can interpret $\langle e_2\rangle$ and
$\langle\fb2\rangle$ as two independent sets of isoscalar factors; hence
$\langle\gb2\rangle$ can be written as a linear combinations of the two sets
of isoscalars. Once again, the argument does not apply to the $i=1$ case; and
for $i=0$, the three sets of matrix elements $\langle e_0\rangle, \langle
\fb0 \rangle$ and $\langle \gb0\rangle$ are proportional to each other as the
operators are all $U(10)$ scalars.  Despite these differences between the
$i=0,1$ and 2 cases, they display the same relation as given
in~\eref{eq:WEefg3}.  We will explain the above relations in a more elegant
way in the next few sections.  Relations~\eref{eq:WEfg1} and~\eref{eq:WEfg2}
will be explained in section~\ref{sec:dir-ex}, and relations~\eref{eq:WEefg1}
-~\eref{eq:WEefg3} in section~\ref{sec:isopure} making use of the isospin
structure.

Before we go on any further, let us pause and reflect on what we have
achieved so far.  We are trying to find all the matrix elements of the nine
operators $e_i$, \fb i and \gb i for the configuration $d^2d'$. The matrix
elements $\langle e_i \rangle$ are rather well known, as they come strictly
from the $d^2$ matrix elements (see Condon and Odabasi, 1980).  Now with
equations~\eref{eq:WEfg1},~\eref{eq:WEfg2} and~\eref{eq:WEefg3}, the \gb i
matrix elements can be easily obtained from those of \fb i and
$e_i$.  Furthermore, with equations~\eref{eq:WEefg1} and~\eref{eq:WEefg2},
the matrix elements $\langle\urs{111}{10}\beta |\fb i|\urs{111}{10}\gamma
\rangle$ and $\langle\urs{111}{10}\beta |\fb i|\urs{21}{10}\gamma\rangle$
are readily obtained from the $e_i$ matrix elements; we are only left with
$\langle\urs{21}{10}\beta |\fb i|\urs{21}{10}\gamma\rangle$ to work on. Is
there an easy way to find these matrix elements? By observation, we find
that the matrix elements of $\eh i \equiv e_i + \fb i$ are surprisingly
simple. For $i=0$, the sum $e_0 + \fb0$ is always 3; for $i=1$, the sum
$e_1+\fb1$ is diagonal, and the value depends only on the $U(5)$ and
$SO(5)$ irreps, but not the $U(10)$ label nor the spin and orbital ranks.
For the case $i=2$, the sum is almost diagonal, with the three exceptions
\beann
\langle d^2d'\ {}^{[T]}U[21](21)^{[S]}D|e_2 + \fb2|
d^2d'\ {}^{[T]}U[21](10)^{[S]}D\rangle &=& 6\surd21 \ .
\eeann
The diagonal values once again do not depend on the $U(10)$ irrep and the
spin, although they do depend on the orbital rank.  These properties, which
will be addressed in section~\ref{sec:opsum}, greatly simplify our task
of finding the \fb i matrix elements.  With these simple relations, (almost)
all the matrix elements can be related to the known $e_i$ matrix elements.
We will elaborate this further in section~\ref{sec:opsum}.

\subsection{The operators \eb i and \et i with simpler $U(10)$
transformation properties\label{sec:dir-ex}}

As mentioned before, the ten creation operators $d^\dag_\mu$ belong to the
irrep \urs1{10} of $U(10)$;
so do the ten components $d'^\dag_\mu$.  When we form products from the two
independent sets, we can get \urs2{10} or \urs{11}{10}.  From the knowledge
of unitary groups and their representations, we know that \urs2{10}
corresponds to the symmetric product, while \urs{11}{10} the antisymmetric
one.  More precisely, $(d^\dag_\mu d'^\dag_\nu + d'^\dag_\mu d^\dag_\nu)$ and
$(d_\mu d'_\nu + d'_\mu d_\nu)$ have 45 components each; they are the
antisymmetric products and belong \urs{11}{10} and \ursb{11}{10} respectively.
The other combination $(d^\dag_\mu d'^\dag_\nu - d'^\dag_\mu d^\dag_\nu)$
belongs to \urs2{10}, has 55 components; and $(d_\mu d'_\nu - d'_\mu d_\nu)$
belongs to \ursb2{10}. Of course, the symmetric product \urs2{10} is
identically zero for two equivalent electrons.

In terms of creation and annihilation operators, \fb i and \gb i can be
written as
\bea
\fb i&=\sum_k2a_{ik} : \bi v^{(k)}_{dd}\cdot\bi v^{(k)}_{d'd'} : \ \sim \sum
C_{\mu \nu} C_{\eta \xi} d^\dag_\mu d'^\dag_\eta d'_\xi d_\nu \nonumber \\
\gb i&=\sum_k2a_{ik} : \bi v^{(k)}_{dd'}\cdot\bi v^{(k)}_{d'd} : \ \sim \sum
C_{\mu \nu} C_{\eta \xi} d^\dag_\mu d'^\dag_\eta d_\xi d'_\nu
\label{eq:fbgb} \eea
where $a_{ik}$ are the corresponding coefficients as given in
equations~\eref{eq:e0} to~\eref{eq:e2}, and $C_{\mu \nu}$ are the appropriate
Clebsch-Gordan (CG) coefficients.  They are invariant under the simultaneous
interchange of $d^\dag \mbox{ with } d'^\dag$ and $d$ with $d'$.  But they
are neither symmetric nor
antisymmetric with respect to the interchange of the creation operators alone.
In fact, if we interchange $d^\dag$ and $d'^\dag$ (or $d$ and $d'$) alone,
\fb i becomes $\gb i$, and vice versa.
If we take the combination
\beann
\eb i = \frac12 (\fb i + \gb i) \sim \sum C_{\mu \nu}
C_{\eta \xi} d^\dag_\mu d'^\dag_\eta (d'_\xi d_\nu + d_\xi d'_\nu)
\eeann
we first note that the annihilation operators within the parenthesis transform
as the antisymmetric
product \ursb{11}{10}.  With a little algebraic manipulation, we can obtain
a similar expression with the creation operators grouped as a \urs{11}{10}
product.  This tells us the new operators \eb i belong to \uro{11}{10} of
$U(10)$.  Similarly, $\et i = \frac12 (\fb i - \gb i)$ is the symmetric
product, and belongs to \uro2{10}.
There are flaws in the above argument; namely, we ignored the scalar parts.
The arguments above do not exclude the possibility that the operators can be
$U(10)$ scalars.  As we already know, \fb0 and \gb0 are both $U(10)$ scalars,
so the sum or difference (\eb0 or \et0) must also be $U(10)$ scalars; although
their expressions look as if they belong to \uro{11}{10} and \uro2{10}
respectively.  Similarly, \eb1 and \et1 can contain a scalar part and, in
fact, they do.  The transformation properties of these new
operators \eb i and \et i are summarized in table~\ref{tb:oplabel}.
In terms of these six new operators, the perturbative Hamiltonian
in ~\eref{eq:h1} becomes
\bea
H_1 & = & \sum_i e_i E^i + \fb i \Fb i + \gb i \Gb i \nonumber \\
& = & \sum_i e_i E^i + \frac{\fb i+ \gb i}2 (\Fb i + \Gb i)
+ \frac{\fb i-\gb i}2 (\Fb i - \Gb i) \nonumber \\
& \equiv & \sum_i e_i E^i + \eb i \Eb i + \et i \Et i \ .
\label{eq:h1new} \eea

\input tbnewmat.tex

From a group-theoretical point of view, it is more convenient to use these
new operators, \eb i and \et i, with simpler transformation properties,
rather than the distinct direct and exchange operators, \fb i and \gb i.
Notice that the operators \eb i have the same transformation properties as
$e_i$.  The matrix elements of \eb i and \et i are given in
table~\ref{tb:newmatrix}, in which we copied down the $e_i$ matrix elements
from table~\ref{tb:matrix} to make it complete. These new operators transform
irreducibly in $U(10)$, makes applying the Wigner-Eckart theorem as well
as the selection rules on the $U(10)$ level so much easier and more elegant.
For example, equations~\eref{eq:WEfg1} and~\eref{eq:WEfg2} are now written as
\numparts\bea
\langle\urs{111}{10}\beta |\et i|\urs{111}{10}\gamma\rangle & = & 0
\label{eq:WEet1} \\
\langle\urs{111}{10}\beta |\et i|\urs{21}{10}\gamma\rangle & = & 0
\label{eq:WEet2} \ .
\eea\endnumparts
From the Kronecker products given in table~\ref{tb:ucross}, neither
$\ursb{111}{10} \times \urs{111}{10}$ nor $\ursb{111}{10} \times \urs{21}{10}$
contains \uro2{10}.  Therefore, equation~\eref{eq:WEet2} as well as the $i=2$
case in~\eref{eq:WEet1} come from the $U(10)$ group selection rule. The
matrix elements $\langle \et0 \rangle = 0$ in~\eref{eq:WEet1} can be obtained
using the derived eigenvalue expression for \et0 in section~\ref{sec:eigen}
after we introduce the isospin structure.  And $\langle\et1\rangle=0$
in~\eref{eq:WEet1} can be verified using the results in
section~\ref{sec:opsum}.  Most of the Wigner-Eckart relations we had before
are now re-interpreted as the selection rule on the group $U(10)$.  This
also explains why the proportionality constants are the same for all $i$,
and why they are 1.

The other three new operators \eb i have identical group labels with the
corresponding $e_i$ operators.  Using the Wigner-Eckart theorem, we find
\numparts\bea
\langle \urs{111}{10}\beta |\eb i|\urs{111}{10}\gamma\rangle & = & 2
\langle \urs{111}{10}\beta |e_i|\urs{111}{10}\gamma\rangle \label{eq:WEeb1} \\
\langle \urs{111}{10}\beta |\eb i|\urs{21}{10}\gamma\rangle & = & -
\langle \urs{111}{10}\beta |e_i|\urs{21}{10}\gamma\rangle \label{eq:WEeb2} \\
\langle \urs{21}{10}\beta |\eb i|\urs{21}{10}\gamma\rangle & = & \frac12
\langle \urs{21}{10}\beta |e_i|\urs{21}{10}\gamma\rangle \label{eq:WEeb3}
\eea\endnumparts
which are the same as equations~\eref{eq:WEefg1} to~\eref{eq:WEefg3}.  Once
again, the $i=1$ cases in~\eref{eq:WEeb1} and~\eref{eq:WEeb3} do not come
from the Wigner-Eckart theorem.  In a sense, we are just rewriting those
three equations without adding much
understanding to the problem we are facing.  But introducing these new
operators \eb i and \et i is a very crucial step towards the full
understanding of the underlying group structure to the problem on hand.

\input ucross.tex

\section{Isospin and the group $U(20)$ \label{sec:iso}}

\v Simonis \etal (1984) first introduced the idea of isospin into the theory
of atomic spectroscopy.  The idea, borrowed from nuclear spectroscopy, is to
treat two electrons with different principal quantum numbers $n$ as two states
($\uparrow$ or $\downarrow$) of a generic $d$-electron in the isospin space.
This gives an elegant explanation to the simple relations observed in the
previous sections, as well as providing new results such as the closed-form
expressions for the scalar operators.  Furthermore, putting $d$ and $d'$
electrons on the same footing lets us conveniently switch between
configurations $d^3$, $d^2d'$ and so forth; hence CI
can be taken into account in a natural manner.

\subsection{Motivation}
In an attempt to exploit the orthogonality relation (Judd and Leavitt, 1986
and Judd \etal 1982) between the nine operators we are looking at (either the
set $e_i$, \fb i, \gb i or $e_i$, \eb i and \et i), it becomes clear that the
situation we are now facing is very different from the equivalent-electron
case.  In the latter one, the operators in question ($e_i$) belong solely to
a single irrep \uro{11}{10} of $U(10)$, with the exception of a complete
scalar operator, $e_0$ (see table~\ref{tb:oplabel}). The three operators $e_0,
e'_1= e_1 -\frac79e_0$ and $e_2$ are orthogonal to each other. Orthogonality
requires summing over all states with the irrep \urs{1^N}{10} (all the states
in $d^3$, for instance).  But now, the states in $d^2d'$ belong to two
different irreps \urs{111}{10} and \urs{21}{10}.  Another very important
aspect about orthogonality is, operators with different transformation
properties are necessarily orthogonal.  For this reason, the $i=2$ operators
are orthogonal to the $i=0,1$ operators, as they transform as (22) and
(00) in $SO(5)$ respectively.  Similarly, \et2 is orthogonal to $e_2$ and
\eb2 as they have different $U(10)$ irreps.  But $e_i$ and \eb i, as well
as \fb i and \gb i, have the same transformation properties.  Can we find a
higher symmetry group to distinguish them?  We also like to have a group such
that all the states in $d^2d'$ belong to a single irrep.

\subsection{The group $U(20)$, its generators and its
subgroup \label{sec:isogroup}}

The 100 generators of the group $U(10)$ in terms of the creation and
annihilation operators are $d^\dag_\mu d_\nu+d'^\dag_\mu d'_\nu$. They can be
re-expressed as $(\bi{d^\dag d})^{(SL)}+(\bi{d'^\dag d'})^{(SL)}$, where
$S=0,1$ and $L=0,\cdots,4$.  This group transforms the states in the
configuration $d^Nd'^{N'}$ among themselves.  If we drop the scalar operators
for which $S=L=0$, we are left with the $SU(10)$ generators. The other
subgroups that we use, $SO_S(3)$ and $U(5)$, are generated by
$(\bi{d^\dag d})^{(10)} + (\bi{d'^\dag d'})^{(10)}$ and
$(\bi{d^\dag d})^{(0L)} + (\bi{d'^\dag d'})^{(0L)}$ respectively.

One can however consider a more general transformation group. If we take the
20 one-electron states ($d^\dag_\mu$ or $d'^\dag_\mu$ acting on the vacuum
$|0\rangle$) as the basis vector space, the most general transformation
among them constitute the unitary group $U(20)$, which has 400 generators of
the form $d^\dag_\mu d_\nu, d'^\dag_\mu d'_\nu, d'^\dag_\mu d_\nu$ and
$d^\dag_\mu d'_\nu$.  These operators preserve electron number, but they
connect states in different configurations, like $d^2d'$ with $d^3$ and so
forth.  The nine operators ($e_i$, \eb i and \et i) now belong to a single
irrep \uro{11}{20} of this higher group plus the scalar \urs0{20}.

A more elegant formulation is given below. Following \v Simonis \etal (1984)
and Kaniauskas \etal (1987), we introduce the operators
$d^{(t\ s\ l)}_{m_tm_sm_l}$ where $t=\frac12$ is the isospin rank, with
$d^{(t\ s\ l)}_{\frac12m_sm_l}=d^{(s\ l)}_{m_sm_l}$,
$d^{(t\ s\ l)}_{-\frac12m_sm_l}=d'^{(s\ l)}_{m_sm_l}$,
$d^{\dag(t\ s\ l)}_{\frac12m_sm_l} = d^{\dag(s\ l)}_{m_sm_l}$ and
$d^{\dag(\ t\ s\ l)}_{-\frac12m_sm_l} = d'^{\dag(s\ l)}_{\ m_sm_l}$.  In this
notation, the 400 generators of $U(20)$ can be written as $w^{(TSL)} =
(\bi{d^\dag d})^{(TSL)}$, where $T,S = 0,1$ and $L=0,\ldots,4$. One can
immediate see that the operators $(\bi{d^\dag d})^{(TSL)}$ with isospin
$T=0$ are simply the $U(10)$ operators described above; and those for which
$T=S=0$ correspond to the $U(5)$ generators.  The three components of
$-\sqrt{10}(\bi{d^\dag d})^{(100)}$
are recognized as the pure isospin generators
$T_+, T_0$ and $T_-$.  In terms of creation and annihilation operators,
\bea
T_+ = d^\dag_\mu d'_\mu \nonumber \\
T_- = d'^\dag_\mu d_\mu \nonumber \\
T_0 = \frac12 ( d^\dag_\mu d_\mu - d'^\dag_\mu d'_\mu ) \ .
\label{eq:isospin} \eea

Obviously, the group $U(20)$ contains $SO_T(3) \times U(10)$ as a formal
subgroup. The subgroup structure $U(20) \rightarrow SO_T(3) \times U(10)$ goes
exactly the same way as $U(10) \rightarrow SO_S(3) \times U(5)$. Augmenting
equation~\eref{eq:subgp}, we can now use the full chain of subgroups
\bea\fl
U(20) \supset SO_T(3) \times U(10) \supset SO_T(3) \times SO_S(3) \times
U(5) \nonumber \\
\supset SO_T(3) \times SO_S(3) \times SO(5) \supset SO_T(3) \times SO_S(3)
\times SO_L(3)
\label{eq:subgp20}\eea

\subsection{Isospin ranks for states and operators}

Since fermions obey the Pauli exclusion principle, no two electrons occupy
the same quantum state (spin and orbital quantum numbers $m_s, m_l$, and
isospin $z$-component $m_t$, which is the same as principal quantum number $n$
in a certain sense). Therefore, any state with $r$ electrons ($d$ or $d'$)
belongs to \urs{1^r}{20} of $U(20)$.  Using the branching rule given in
table~\ref{tb:ugpbr}, $\urs{1^r}{20} \rightarrow {}\surs{r+1}{1^r}{10} +
{}\surs{r-1}{21^{(r-2)}}{10} + \cdots$, we can quickly write down the isospin
rank for all the states. Using the same notation as for regular spin, we put
the multiplicity $[T] = 2T+1$ as a superscript preceding the $U(10)$ label.
They are given in table~\ref{tb:states}.  Notice that the isospin ranks are
uniquely determined by the $U(10)$ labels for atomic states. (This is not true
for operators, however.) As for the isospin $z$-component $M_T$, it is simply
given by the configuration type.  From equation~\eref{eq:isospin}, we can see
that for the configuration $d^Nd'^{N'}$, $M_T = \frac12(N-N')$.

Let us now look at the isospin rank of the Coulomb operators.  The operators
we have considered so far, which do not include CI,
all have $M_T = 0$.  This can be verified by evaluating the commutator
$[T_0,e_0]=0$ for example (using equations~\eref{eq:isospin}
and~\eref{eq:e0create}).  From now on, we will omit the commutator
brackets, knowing that the adjoint action (group generators acting on
operators) always means taking the commutator; and the above equation
reads $T_0e_0 =0$.  Since we know that all Coulomb operators belong to
\uro{11}{20} and \urs0{20} of $U(20)$, the branching rules $U(20) \rightarrow
SO_T(3) \times U(10)$ (from table~\ref{tb:ugpbr}) will give us the possible
ranks for the operators.  The branching rules uniquely determine the isospin
rank of the operator \et2, which belongs to \suro12{10}.  In fact, we can
easily show that \et0 and \et1 are also isospin scalars as well.  From
equation~\eref{eq:fbgb}, we have $\et i \sim \sum C_{\mu \nu} C_{\eta \xi}
(d^\dag_\mu d'^\dag_\eta d'_\xi d_\nu -d^\dag_\mu d'^\dag_\eta
d_\xi d'_\nu)$.  Together with equation~\eref{eq:isospin}, we find
$T_+\et i=0$ for all $i$.  So, they are all isospin scalars.

For the other six operators $e_i$ and \eb i, the possible isospin ranks as
given by the branching rules are \surs{1,5}0{10} and \suro{1,3,5}{11}{10}.
The immediate question that comes to mind is whether we can separate the
various spin rank components from these operators. Repeatedly applying the
raising operator $T_+$ to them should eliminate the lower rank components.
One of the results that we find is $T_-T_+^2 \eb i$ is proportional to $T_+
\eb i$.  This tells us that the operators \eb i do not have a rank $1$
component.  So, they belong only to \surs{1,5}0{10} and \suro{1,5}{11}{10}.
The results are summarized in table~\ref{tb:oplabel}.

\subsection{Operators with pure isospin rank\label{sec:isopure}}

We will now continue our effort to separate the \eb i (and $e_i$) operators
into various isospin components.  The rank 2 component of \eb i is given by
$\frac1{24} T_-^2 T_+^2 \eb i$; as we already know there is no spin 1
component, the remaining piece must be the isospin scalar part.
From equation~\eref{eq:fbgb},
\bea
\eb i&=\frac12(\fb i+\gb i)=\sum_k a_{ik}\{ : \bi v^{(k)}_{dd}\cdot\bi
v^{(k)}_{d'd'}: + : \bi v^{(k)}_{dd'}\cdot\bi v^{(k)}_{d'd} : \} \nonumber \\
& \sim \sum C_{\mu \nu} C_{\eta \xi} (d^\dag_\mu
d'^\dag_\eta d'_\xi d_\nu + d^\dag_\mu d'^\dag_\eta d_\xi d'_\nu ) \ .
\label{eq:eb}\eea
Let us take the simplest one, $\eb0 = \frac12 ( d^\dag_\mu d'^\dag_\nu
d'_\nu d_\mu + d^\dag_\mu d'^\dag_\nu d_\nu d'_\mu )$, as an example.  After
taking four consecutive commutations on \eb0, we find
\bea
T_-^2T_+^2 \eb0 = 16 \eb0 - 4 ( d^\dag_\mu d^\dag_\nu d_\nu
d_\mu + d'^\dag_\mu d'^\dag_\nu d'_\nu d'_\mu) \ .
\label{eq:t4e0}\eea
The first term within the parenthesis is recognized as $2e_0$ (see
equation~\eref{eq:e0create}).  The second one is the analogous term for the
$d'$-electron, which does not
contribute in our problem as we have only one $d'$-electron.  We would like
to absorb the second term into our definition of $e_0$, and
$e_i$ in general, and so we write
\bea
e_i &= \sum_k a_{ik} \{ : \bi v^{(k)}_{dd}\cdot\bi v^{(k)}_{dd} : +
 : \bi v^{(k)}_{d'd'}\cdot\bi v^{(k)}_{d'd'} : \} \nonumber \\
& \sim \sum C_{\mu \nu} C_{\eta \xi} (d^\dag_\mu d^\dag_\eta d_\xi
d_\nu + d'^\dag_\mu d'^\dag_\eta d'_\xi d'_\nu ) \ .
\label{eq:e} \eea
Comparing this result to equation~\eref{eq:eb}, we see that $e_i$ and \eb i
are almost the same; after all, they have the same group transformation
properties.  Of course, in the complete analysis of configurations with at
least two $d'$ and two $d$-electrons, one should include three new operators
of the kind $d^\dag d^\dag d d - d'^\dag d'^\dag d' d'$ to account for the
interactions between the two $d'$-electrons.

With our modified definition of $e_0$, the right hand side of
equation~\eref{eq:t4e0} becomes $16\eb0 -8e_0$, and is a pure isospin rank 2
operator.  Repeating the exercise, and keeping track of the CG coefficients
carefully, one finds that this is true for any $i$.  The combination
$2\eb i - e_i$ has isospin rank 2.  We can now separate \eb i as
$\{\frac13 (2\eb i - e_i) + \frac13(\eb i + e_i) \}$, with the first term
purely isospin rank 2, and the second one an isospin scalar.  One can also
verify $T_+ (\eb i + e_i) =0$ easily.

In principle, we can re-tabulate the matrix elements using these six new
operators $\eb i + e_i$ and $2\eb i - e_i$  that have simpler transformation
properties in place of $e_i$ and \eb i.  However, we also feel that the
relatively simple expressions of $e_i$ and \eb i (as in
equations~\eref{eq:eb} and~\eref{eq:e}) have their merits. Moreover,
the $e_i$ matrix elements are easily obtained from the $d^2$ matrix
elements. We would rather leave them untouched.  The new combinations, with
pure isospin ranks, can nevertheless give us a lot of useful information
and insight into the problem we are facing.

Recall that we have found some simple proportionality relations in
equations~\eref{eq:WEeb1}--~\eref{eq:WEeb3}.  The underlying reasons are as
follows.  The combination $2\eb i - e_i$, with isospin rank 2, must have
vanishing matrix elements if sandwiched between a pair of \surs2{21}{10}
states, since two $T=\frac12$ states cannot be stretched to $T=2$.  Therefore,
$\langle\eb i\rangle = \frac12 \langle e_i\rangle$ for these matrix elements,
and thus we arrive at relation~\eref{eq:WEeb3}.  And the other combination
$\eb i + e_i$, being an isospin scalar, must be diagonal in isospin space.
So, it vanishes between a pair of \surs4{111}{10}
and \surs2{21}{10} states.  In other words, $\langle\eb i\rangle = -
\langle e_i\rangle$, as in equation~\eref{eq:WEeb2}.  From our new
perspective, the two Wigner-Eckart relations we had before are now
interpreted as selection rules on the isospin group.

To explain relation~\eref{eq:WEeb1} requires a more elaborate analysis.
First notice that the \surs4{111}{10} states also appear in the configuration
$d^3$.  In fact, they are almost the same state, with the only difference in
isospin $z$-component. Recall that states in $d^3$ have isospin
$M_T=\frac32$, while the $d^2d'$ states
have $M_T=\frac12$.  It is well-known that matrix elements can be factored as
\bea
\langle\alpha T M_T | U^{(k)}_q|\alpha' T' M'_T\rangle = (-1)^{T-M_T}
\tj Tk{T'}{-M_T}q{M'_T} (\alpha T\|U^{(k)}\|\alpha' T')
\label{eq:reduce}\eea
where the last factor is the reduced matrix element.
So, for the isospin scalar operator $\eb i + e_i$, we have
\bea\fl
%\hspace{-2em}
\frac{\langle d^2d' \ \surs4{111}{10}\beta| \eb i + e_i |
d^2d' \ \surs4{111}{10}\gamma\rangle}
{\langle d^3 \ \surs4{111}{10}\beta| \eb i + e_i |
d^3 \ \surs4{111}{10}\gamma\rangle}
= -\tj{\frac32}0{\frac32}{-\frac12}0{\frac12}
\tj{\frac32}0{\frac32}{-\frac32}0{\frac32}^{-1} = 1 \ ,
\label{eq:ratio1}\eea
while for the isospin rank 2 operator $2\eb i - e_i$, we have
\bea\fl
%\hspace{-2em}
\frac{\langle d^2d' \ \surs4{111}{10} \beta| 2\eb i - e_i |
d^2d' \ \surs4{111}{10}\gamma\rangle}
{ \langle d^3 \ \surs4{111}{10} \beta| 2\eb i - e_i |
d^3 \ \surs4{111}{10} \gamma\rangle}
= -\tj{\frac32}2{\frac32}{-\frac12}0{\frac12}
\tj{\frac32}2{\frac32}{-\frac32}0{\frac32}^{-1} = -1 \ .
\label{eq:ratio2}\eea
But $\langle d^3 \ \surs4{111}{10}\beta| \eb i|d^3 \
\surs4{111}{10}\gamma\rangle = 0$ in $d^3$ since there is no $d'$ electron.
Therefore from these two equations, we can solve for $\langle d^3 \
\surs4{111}{10}\beta|e_i|d^3 \ \surs4{111}{10}\gamma\rangle$ and $\langle
d^2d' \ \surs4{111}{10}\beta|\eb i|d^2d' \ \surs4{111}{10}\gamma\rangle$ in
terms of $\langle d^2d' \ \surs4{111}{10}\beta|e_i|d^2d' \
\surs4{111}{10}\gamma\rangle$ which are assumed known; and the answer is
\numparts\bea
\langle d^2d' \ \urs{111}{10}\beta |\eb i|d^2d' \ \urs{111}{10}\gamma\rangle
& = & 2 \langle d^2d' \ \urs{111}{10}\beta |e_i|d^2d' \
\urs{111}{10}\gamma\rangle \\
\langle d^3 \ \urs{111}{10}\beta |e_i|d^3 \ \urs{111}{10}\gamma\rangle
& = & 3 \langle d^2d' \ \urs{111}{10}\beta |e_i|d^2d' \
\urs{111}{10}\gamma\rangle \ .  \eea\endnumparts
The first one is of course relation~\eref{eq:WEeb1} that we are aiming at;
the second one is a bonus, that relates the $d^2d'$ matrix elements to the
ones in $d^3$.  Of course, the $d^3$ matrix elements are also available in
the literature (see Nielson and Koster 1963 for instance), so the second
equation can be served as a consistency check.

\subsection{Configuration mixing via Coulomb interaction}

The isospin arguments used above can relate the $d^3$ Coulomb matrix elements
to those in $d^2d'$.  In the same way, we can obtain the
mixed matrix elements between configurations $d^3$ and $d^2d'$.
The first point to note is that the Coulomb operators which connect $d^3$
with $d^2d'$ necessarily have $M_T=1$; so no isospin scalar Coulomb operator
can be responsible for CI.  Let us write
\A2 qi as the isospin rank 2 Coulomb operator, such that
\beann
\A20i &= \sqrt{\frac23} (2\eb i -e_i)\\
&=\sqrt{\frac23}\sum C_{\mu \nu} C_{\eta \xi}( d^\dag_\mu d'^\dag_\eta d'_\xi
d_\nu + d^\dag_\mu d'^\dag_\eta d_\xi d'_\nu -\frac12 d^\dag_\mu d^\dag_\eta
d_\xi d_\nu - \frac12 d'^\dag_\mu d'^\dag_\eta d'_\xi d'_\nu ) \\
\A21i &= \sum C_{\mu \nu} C_{\eta \xi} (d^\dag_\mu d^\dag_\eta d'_\xi d_\nu -
d'^\dag_\mu d^\dag_\eta d'_\xi d'_\nu ) \\
\A22i &=-\sum C_{\mu\nu}C_{\eta \xi}(d^\dag_\mu d^\dag_\eta d'_\xi d'_\nu ) \ .
\eeann
With the notation developed before, the above operators are simply the
components of the tensor operator $10(w^{(10k)} \cdot w^{(10k)})^{(200)}$.
Parallel to equations~\eref{eq:ratio1} and~\eref{eq:ratio2}, we have
\numparts\bea\fl
%\hspace{-3em}
\frac{\langle d^3 \ \surs4{111}{10}\beta|\A21i|d^2d' \ \surs4{111}{10}\gamma
\rangle}{\langle d^2d'\ \surs4{111}{10}\beta|\A20i|d^2d'\ \surs4{111}{10}\gamma
\rangle} &= -\tj{\frac32}2{\frac32}{-\frac32}1{\frac12}\tj{\frac32}2{\frac32}
{-\frac12}0{\frac12}^{-1} =\frac1{\surd2} \label{eq:mix1}\\
%\hspace{-3em}
\fl\frac{\langle d^3\ \surs4{111}{10}\beta|\A21i|d^2d'\ \surs2{21}{10}\gamma
\rangle}{\langle d^2d'\ \surs4{111}{10}\beta|\A20i|d^2d'\ \surs2{21}{10}
\gamma\rangle} &= -\tj{\frac32}2{\frac12}{-\frac32}1{\frac12}
\tj{\frac32}2{\frac12}{-\frac12}0{\frac12}^{-1} = \surd2 \ .
\label{eq:mix2}\eea\endnumparts

The other possible configuration mixing Coulomb operators ($M_T=1$) are given
by $(w^{(10k)} \cdot w^{(10k)})^{(100)}$ and $\A1{}i \sim (w^{(00k)} \cdot
w^{(10k)})^{(100)}$. The first one is in fact identically zero when normal
ordered. The second one with $M_T=0$ are the isospin rank one operators
($i=0,1,2$) that we briefly mentioned in section~\ref{sec:isopure}. They
are of the form $d^\dag d^\dag dd - d'^\dag d'^\dag d'd'$, as opposed to the
operators $e_i$ in~\eref{eq:e}. In the configurations with no more than one
$d'$ electron, they have the same matrix elements as the $e_i$.  So in this
manner, we can determine all the configuration-interaction Coulomb matrix
elements between $d^3$ and $d^2d'$ from the known ones in $d^2d'$.  We can
similarly write down the other matrix elements between $d^3$ and $dd'^2$,
$d^2d'$ and $dd'^2$ and so forth.

Before we close this section, we should mention the relevance of Brillouin's
theorem (see Bauche and Klapisch 1972, and Godefroid \etal 1987) in our
analysis. The theorem says, the Hartree-Fock (HF) solution $\Psi_{HF}$ of the
configuration $l^N$ has vanishing matrix elements with a class of states in
$l^{N-1}l'$. They are the states in which the $l'$ electron is coupled to the
$l^{N-1}$ state via the ordinary fractional parentage coefficients as
in $l^N$.  That is to say, states in $l^{N-1}l'$ and $l^N$ with the same
angular form will not mix via Coulomb interaction, if the HF solution for the
$l^N$ configuration is used.  In our case, $\langle d^3 \ \surs4{111}{10}
\beta| \mbox{Coulomb} | d^2d' \ \surs4{111}{10} \beta\rangle$ will have a
vanishing radial integral.  That does not make our work less useful, however.
In order to apply Brillouin's theorem in our situation, one has to first find
out the HF solution $\Psi_{HF}$ for each single $LS$ term in the configuration
$l^N$, which might be straightforward, but definitely not easy.  Only then,
equation~\eref{eq:mix1} will rendered irrelevant, as Brillouin's theorem
predicts zero radial integrals in those cases.  All other results are
otherwise unaffected.

\subsection{Eigenvalues for the scalar operators\label{sec:eigen}}

\input casimir.tex

From table~\ref{tb:oplabel}, there are a few complete scalar operators. We
can now find closed-form expressions for each $U(10)$ scalar operator; that
is, the $i=0$ operators.  Let us start from the simplest operator $e_0$. From
equations~\eref{eq:e0create} and~\eref{eq:e}, we can see that the eigenvalue
for $e_0$ is simply given by $\frac12 N(N-1) + \frac12 N'(N'-1)$.  For \fb0,
from equations~\eref{eq:f0create}, it is equals to $NN'$. For the next one,
\gb0, from equations~\eref{eq:g0create} and~\eref{eq:isospin}, we find
$\gb0 = \ :\! T_+T_-\! :\ = T_+T_- - N$.
None of these three operators $e_0$, \fb0 and \gb0 is an isospin scalar. The
other two operators \eb0 and \et0 are just linear combinations of \fb0 and
\gb0. The results are summarized in the table~\ref{tb:casimir}. Note that the
operator \et0 is an isospin scalar, so is the combination $\eb0 + e_0$. Their
expressions from table~\ref{tb:casimir} certainly verify that fact; as $T^2$
is a scalar in $SO_T(3) \times U(10)$ and the total electron number $N_T=N+
N'$ is, in fact, a scalar in $U(20)$ and hence scalar in all its subgroups.

\section{The operator sum $\bi{\widehat e_i = e_i+ \overline f_i = e_i +
\overline e_i + \widetilde e_i}$ \label{sec:opsum}}

Recall that at the end of section~\ref{sec:selection}, we mentioned a simple
result on the operators $\eh i \equiv e_i + \fb i$.  In terms of the new
operators, the sum is $\eh i \equiv e_i + \eb i + \et i$; whose matrix
elements are included in table~\ref{tb:newmatrix}.  As a brief summary of our
work up to this point, the $e_i$ matrix elements of $d^2d'$ are obtained
easily from the ones in the configuration $d^2$.  The \eb i matrix elements
can be related to $e_i$ by just three constants using the isospin structure
as in~\eref{eq:WEeb1} -~\eref{eq:WEeb3}.  For the last operator \et i,
selection rules on $SO_T(3) \times U(10)$ give us a lot of vanishing matrix
elements (equations~\eref{eq:WEet1} and~\eref{eq:WEet2}); the rest involving
a pair of \urs{21}{10} states are yet to be found.  A thorough understanding
on the simple result for the operator \eh i will definitely help to
accomplish our plan for finding the remaining matrix elements.

Let us look at the three different cases separately.  For $i=0$, \eh0 always
equals 3 in $d^2d'$.  From table~\ref{tb:casimir}, one can see that the
operator sum equals $\frac12N_T(N_T-1)$, which is scalar in $U(20)$.  Since
all the states belong to the same $U(20)$ multiplet \urs{111}{20}, they must
have the same matrix element.

The next one \eh1 is also diagonal; with matrix elements 0,3,6,7 and 13 only,
which are determined by the $U(5)$ and $SO(5)$ irreps.  To explain this, we
write down the operator in terms of $\bi v^{(k)}=\bi v^{(k)}_{dd}+\bi
v^{(k)}_{d'd'}=2\bi w^{(00k)}$.  Using the extended version of
equations~\eref{eq:e1} (recall that we have added the identical terms
involving the $d'$-electron to the operators $e_i$ in
section~\ref{sec:isopure}), the operator \eh1 takes the simple form
\beann
\frac72:\bi v^{(0)}\cdot\bi v^{(0)}:\ +\ :\bi v^{(2)}\cdot\bi v^{(2)} : \
+ \ :\bi v^{(4)}\cdot\bi v^{(4)} :
\eeann
which is reminiscent of the (quadratic) Casimir operators for the groups
$SU(2l+1)$ and $SO(2l+1)$; they are given by \( {\cal C}(SU_{2l+1})=
{\displaystyle\sum_{k>0}}\bi v^{(k)}\cdot\bi v^{(k)} \) and \( {\cal C}
(SO_{2l+1})={\displaystyle\sum_{odd \ k}}\bi v^{(k)}\cdot\bi v^{(k)}\)
respectively.  For the irrep $[\lambda_1,\lambda_2,\cdots]_{2l+1}$ of
$U(2l+1)$, the eigenvalue of ${\cal C}(SU_{2l+1})$ is (see Judd 1998, \S5.9)
\bea
\sum_{i=1}^{2l+1} \lambda_i(\lambda_i +2+2l-2i) -\frac{n^2}{[l]}
\eea
where $n=\sum \lambda_i$. One can check that $[\lambda_1,\lambda_2,
\cdots]_{2l+1}$ and $[\lambda_1+\alpha,\lambda_2+\alpha,\cdots]_{2l+1}$ both
have the same eigenvalue as they should since they possess the same $SU(2l+1)$
content. The eigenvalue of ${\cal C}(SO_{2l+1})$ on the irrep
$(w_1,w_2,\cdots)$ of $SO(2l+1)$ is
\bea
\frac12 \sum_{i=1}^l w_i(w_i +1+2l-2i) \ .
\eea
We can remove the normal ordering using the relation $:\bi v^{(k)}\cdot\bi
v^{(k)}: \ =\bi v^{(k)}\cdot\bi v^{(k)}-\frac{[k]}{[l]}N_T$, where $l=2$ in
our case. Putting these together with $\bi v^{(0)}\cdot\bi v^{(0)}=N_T^2 /
[l]$, we find
\beann
\eh1 & = \frac7{10} :N_T^2: + :{\cal C}(SU_5) - {\cal C}(SO_5): \\
& = \frac7{10}N_T^2 - \frac72N_T + {\cal C}(SU_5) - {\cal C}(SO_5) \ .
\eeann
This explains why \eh1 is diagonal, and gives the correct eigenvalues. As we
know $\langle e_1 \rangle$ and $\langle \eb1 \rangle$ already, this simple
result can also be used to determine $\langle \et1 \rangle = 0$ in
equation~\eref{eq:WEet1}, where vanishing matrix elements cannot be explained
by $U(10)$ selection rule.

Finally, the operator \eh2 is almost diagonal, with the three exceptions
\beann
\langle d^2d'\ {}^{[T]}U[21](21)^{[S]}D|\eh2|
d^2d'\ {}^{[T]}U[21](10)^{[S]}D\rangle &=& 6\surd21 \ .
\eeann
Since the operators $e_2+\eb2$ and \et2 are both isospin scalars, clearly the
sum must also be diagonal in isospin space.  With a little hindsight, we can
further conclude that it is diagonal in the $U(5)$ space as well; so that the
\urs{21}5 states will not mix with \urs{111}5 or \urs35 states\footnote{The
operator is not a $U(5)$ scalar however; $U(5)$ scalars do not give (22) of
$SO(5)$.}. The above three exceptions are the only possible off-diagonal
entries we can have. Furthermore, the matrix elements for a set of $L$ states
do not depend on the spin, isospin or $U(10)$ irreps; they depend only on their
respective $SO(5)$ irreps. All these can be explained using the spin-isospin
supermultiplet group $SU(4)$ of Wigner (1937).

To explain these, let us first introduce an alternative branching scheme
$U(20) \supset SO_S(3) \times U'(10) \supset SO_S(3) \times SO_T(3) \times
U(5)$ as oppose to the one in~\eref{eq:subgp20}.  The $U'(10)$ group
acts on the isospin-orbital space, which is analogous but different from the
spin-orbital $U(10)$ group that we had before; however, the $U(5)$ group is
the same in both schemes.  To
better display the symmetry, in the original scheme, we label a state with
its spin superscript $[S]$ put before the $U(5)$ label.  For example, we
now write the first of the doublet $F$ states and the last of the quartet $F$
states as $|\surs4{111}{10}\surs2{21}5(21)F \rangle$ and
$|\surs2{21}{10}\surs4{21}5(21)F \rangle$; which become
$|^2[21]'_{10}\surs4{21}5(21)F \rangle$ and
$|^4[111]'_{10}\surs2{21}5(21)F \rangle$ respectively in the new scheme. With
a mere spin-isospin exchange, the doublet $F$ state switches with the quartet
$F$ state; on the other hand, the operator \eh2 is invariant.  This explain
why the two states have the same matrix element.  But there is yet another
doublet $F$ state, $|\surs2{21}{10}\surs2{21}5(21)F \rangle$, that has the
same diagonal matrix element. To accommodate this, we should promote the idea
of spin-isospin exchange to a more general transformation in the spin-isospin
space.  This leads us to introduce the $SU(4)$ supermultiplet group due to
Wigner (1937).  The three $F$ states now fall into a single supermultiplet
$\urs{21}4 \otimes\urs{21}5$ of $SU(4) \times U(5)\subset U(20)$\footnote{We
use the simpler $U(4)$ irreps rather than $SU(4)$ irreps throughout the paper,
which should not affect the validity of our arguments and results.}.  The
other possible multiplets in $d^2d'$ are $\urs{111}4 \otimes \urs35$ and
$\urs34 \otimes \urs{111}5$.  The operator \eh2, being a $SU(4)$ scalar, is
diagonal in the $SU(4)$ space; which appears as if it is diagonal in the
orbital $U(5)$ space since each $U(5)$ irrep is paired with a unique $SU(4)$
irrep in the problem on hand.  This also explain why the matrix elements are
independent of spin or isospin ranks, as they belong to the same multiplet
in $SU(4) \times SO(5)$.

As explained before, $\langle d^2d' \surs2{21}{10}\beta |\et2|d^2d'
\surs2{21}{10}\gamma \rangle$ are the only matrix elements for which we have
not yet found a simple way to deduce the values. With this latest result, we
can obtain those matrix elements easily if $\beta, \gamma$ do not contain the
irreps $\urs35(30)$ of $U(5)$ and $SO(5)$. To fill in the last piece of puzzle,
we make use of the following operator introduced by Judd (1998, p~222):
\beann
\Omega ' =7\bi v^{(1)}\cdot\bi v^{(1)} - 3\bi v^{(3)}\cdot\bi v^{(3)}
= -3{\cal C}(SO_5) + L^2 \ .
\eeann
It transforms as (22) in $SO(5)$, just like the $i=2$ operators; and its
eigenvalue is easy to compute. So for the $|\surs2{21}{10}\urs35(30)^2SGHI
\rangle$ states, since $(30) \times (30)$ contains (22) only once, we can
apply the Wigner-Eckart theorem and find
\beann
\langle ^2[21][3](30)^2L|\eh2| ^2[21][3](30)^2L\rangle
= \langle ^2[21][3](30)^2L|\Omega '| ^2[21][3](30)^2L\rangle \ .
\eeann
In other words, we only need to calculate the matrix element for one of the
four terms $SGHI$ to determine the proportionality constant (which turns out
to be unity). The other three can be determined from the above relation
easily.  Usually, the calculation on the fully stretched state ($^2I$) is
reasonably easy.  In the end, we find that $\langle\,^2I|\fb2|\,^2I\rangle$
is the only matrix element we need to calculate.  All other $d^2d'$ matrix
elements are related to $e_i$ ones via some simple arithmetic relations. This
is a truly surprising result from the present analysis.

\section{Spin-Orbit interaction}
From a group-theoretical point of view, the spin-orbit interaction is much
simpler than the Coulomb interaction.  The interaction Hamiltonian, in second
quantized form, is:
\beann
H_2=\zeta_{dd}\ w^{(11)0}_{dd}+\zeta_{d'd'}\ w^{(11)0}_{d'd'}
\eeann
where the $\zeta_{dd}$ and $\zeta_{d'd'}$ are the corresponding radial
integrals; the symbol $\kappa$ following the ranks $(SL)$ for the coupled
tensor $\bi w^{(SL)\kappa}_{dd} = -(\bi{d^\dag d})^{(SL)\kappa}$ is the rank
to which $S$ and $L$ coupled (see Judd 1967). The radial integral
$\zeta_{dd}$ is related to the classic parameter $\zeta$ of Condon and
Shortley (1953) by $\zeta_{dd} = -\sqrt{15}\zeta$.  In the same manner, we
can include CI into our analysis by introducing the perturbing Hamiltonian
\beann
H_2^{mix} =\zeta_{dd'}\ w^{(11)0}_{dd'}+\zeta_{d'd}\ w^{(11)0}_{d'd} \ .
\eeann
The operators $w^{(11)0}_{dd'}$ and $w^{(11)0}_{d'd}$ are simply
hermitian conjugate to each other. Each of the above four operators belongs
to \uro1{20}, the adjoint representation of $U(20)$; since the nine
components of $\bi w^{(11)}$ are the $SU(20)$ group generators (see
section~\ref{sec:isogroup}), the coupled scalar operator must also belong to
the adjoint representation.  Using the branching rule $U(20) \rightarrow
SO_T(3) \times U(10)$ given in table~\ref{tb:ugpbr}, \uro1{20} gives
\suro{1,3}1{10} + \surs{1,3}0{10}. Only the irreps \suro{1,3}1{10} can
eventually give us a $S=L=1$ ($^3P$) tensor in the ordinary spin and orbital
spaces, to which the spin-orbit operators belong; hence the $U(10)$ scalar
irreps can be dropped.  Then from $U(10) \rightarrow SO_S(3) \times U(5)$,
\uro1{10} gives us $\suro{1,3}15 + \surs{1,3}05$; once again, the $U(5)$
scalar irreps can be dropped.  And this time we only want the spin 1 part, so
\suro315 is the only permissible label for the spin-orbit operators on the
$SO_S(3) \times U(5)$ level.

Finally, on the $SO(5)$ level, using the branching rules $U(5) \rightarrow
SO(5)$, \uro15 gives us $(11)+(20)$ of $SO(5)$, but (20) contains only
$S,D$ and $G$ terms in $SO_L(3)$ while we are looking for a $P$ term.  So,
we can conclude that the spin orbit operators belong to
$\uro1{20}{}\suro{1,3}1{10}\uro15(11)^3P$ in our branching scheme.

\subsection{Relations between various spin-orbit matrix elements on the
$U(10)$ level}
Using the Wigner-Eckart theorem, we can take advantage of the known
spin-orbit interaction matrix elements of the configuration $d^3$ (see
Nielson and Koster 1963, for instance), in which all the states belong to
\surs4{111}{10} of $SO_T(3) \times U(10)$. Since $\ursb{111}{10} \times
\urs{111}{10}$ contains the adjoint irrep \uro1{10} once, on the $U(10)$
level we can write down several direct proportionalities. With the
numerical constants, they run
\numparts\bea\fl
(d^3 \ \surs4{111}{10}\beta\|w^{(11)}_{dd}\|
d^3 \ \surs4{111}{10}\gamma) & =
3 (d^2d'\ \surs4{111}{10}\beta\|w^{(11)}_{d'd'}\|d^2d'\
\surs4{111}{10}\gamma) \label{eq:spin1a} \\
& = \frac32 (d^2d'\ \surs4{111}{10}\beta\|w^{(11)}_{dd}\|d^2d'\
\surs4{111}{10}\gamma) \label{eq:spin1b} \\
& = \surd3 (d^3\ \surs4{111}{10}\beta\|w^{(11)}_{dd'}\|d^2d'\
\surs4{111}{10}\gamma) \ .
\label{eq:spin1c}\eea
Since these involve reduced matrix elements, no components of $\bi w^{(11)}$
or magnetic quantum numbers $M_S$ and $M_L$ appear; hence the extra
information $\kappa =0$ is rendered irrelevant. So, with just three
constants, we can determine all the spin-orbit matrix elements between a pair
of \surs4{111}{10} states in $d^2d'$, as well as the CI matrix elements
between $d^2d'$ and $d^3$.

For matrix elements involving the \surs2{21}{10} states, there are no such
corresponding states in the $d^3$ configuration.  But still, for matrix
elements between the \surs4{111}{10} and \surs2{21}{10} states, since
$\ursb{111}{10} \times \urs{21}{10}$ contains the adjoint only once, we can
relate the reduced matrix elements by direct proportionalities, which turn
out to be
\bea\fl
( d^2d'\ \surs4{111}{10}\beta\|w^{(11)}_{dd}\|d^2d'\ \surs2{21}{10}\gamma)
& = - ( d^2d'\ \surs4{111}{10}\beta\|w^{(11)}_{d'd'}\|
d^2d'\ \surs2{21}{10}\gamma) \label{eq:spin2a} \\
& = -\frac1{\surd3} ( d^3\ \surs4{111}{10}\beta\|w^{(11)}_{dd'}\|d^2d'\
\surs2{21}{10}\gamma)
\label{eq:spin2b}\eea\endnumparts

Lastly, for matrix elements between a pair of \surs2{21}{10} states, the
simple version of the Wigner-Eckart theorem does not apply, since
$\ursb{21}{10} \times \urs{21}{10}$ contains the adjoint \uro1{10} twice. The
above equations are all we can get from the Wigner-Eckart theorem on the
$U(10)$ level.  One direction to proceed is to seek for similar relations on
the $U(5)$ and $SO(5)$ levels, so that most of the remaining matrix elements
can eventually be related to the ones in $d^3$.  We can also look at the use
of isospin in the spin-orbit analysis.

\subsection{The use of isospin in spin-orbit interaction}
In order to take advantage of the underlying isospin structure in the
spin-orbit interaction, we should first decompose the spin-orbit interaction
operators into components with definite isospin ranks.  The four operators
$\bi w^{(11)}_{dd},\bi w^{(11)}_{d'd'},\bi w^{(11)}_{dd'}$ and $\bi
w^{(11)}_{d'd}$ belong to \suro{1,3}1{10} in $SO_T(3) \times U(10)$.  Among
them, $\bi w^{(11)}_{dd'}$ and $\bi w^{(11)}_{d'd}$ has $M_T=1$ and $-1$
respectively; so they must belong solely to \suro31{10}.  For the other two
operators, $\bi w^{(11)}_{dd}$ and $\bi w^{(11)}_{d'd'}$, with $M_T=0$, one
can check that $\bi w^{(11)}_{dd}+\bi w^{(11)}_{d'd'}$ is an isospin scalar
(i.e.\ commutes with $T_{+/-}$); and $\bi w^{(11)}_{dd}-\bi w^{(11)}_{d'd'}$
has an isospin rank 1. More precisely, $-\bi w^{(11)}_{dd'}, \frac1{\surd2}
(\bi w^{(11)}_{dd}-\bi w^{(11)}_{d'd'})$ and $\bi w^{(11)}_{d'd}$ are the
three components $M_T=0,\pm1$ of the operator $3\bi w^{(111)} = 3(\bi
{d^\dag d})^{(111)}$; while $-\sqrt2\bi w^{(011)} = (\bi w^{(11)}_{dd}+\bi
w^{(11)}_{d'd'})$ is the isospin scalar operator.

We can now easily deduce equations~\eref{eq:spin1a} to~\eref{eq:spin2b} from
isospin arguments.  First of all, the scalar operator just mentioned is
diagonal in isospin space, therefore we have
\beann
( d^2d'\ \surs4{111}{10}\beta\|w^{(11)}_{dd} + w^{(11)}_{d'd'}\|
d^2d'\ \surs2{21}{10}\gamma) = 0
\eeann
which proves relation~\eref{eq:spin2a}.  Then, using equation~\eref{eq:reduce}
to remove the $M_T$ dependence of the two matrix elements
\beann
( d^3\ \surs4{111}{10}\beta\|w^{(11)}_{dd'}\|d^2d'\ \surs2{21}{10}\gamma)
\left( = -\frac1{\sqrt6}
(\surs4{111}{10}\beta|||3w^{(111)}||| \surs2{21}{10}\gamma) \right) \\
( d^2d'\ \surs4{111}{10}\beta\|\frac1{\surd2}(w^{(11)}_{dd} -
w^{(11)}_{d'd'})\|d^2d'\ \surs2{21}{10}\gamma)
\eeann
we can easily get relation~\eref{eq:spin2b}.  Similarly, we can express
\beann
( d^3\ \surs4{111}{10}\beta\|\frac1{\surd2}(w^{(11)}_{dd} +
w^{(11)}_{d'd'})\|d^3\ \surs4{111}{10}\gamma) \\
( d^2d'\ \surs4{111}{10}\beta\|\frac1{\surd2}(w^{(11)}_{dd} +
w^{(11)}_{d'd'})\|d^2d'\ \surs4{111}{10}\gamma) \\
( d^3\ \surs4{111}{10}\beta\|\frac1{\surd2}(w^{(11)}_{dd} -
w^{(11)}_{d'd'})\|d^3\ \surs4{111}{10}\gamma) \\
( d^2d'\ \surs4{111}{10}\beta\|\frac1{\surd2}(w^{(11)}_{dd} -
w^{(11)}_{d'd'})\|d^2d'\ \surs4{111}{10}\gamma) \\
( d^3\ \surs4{111}{10}\beta\|w^{(11)}_{dd'}\|d^2d'\ \surs4{111}{10}\gamma)
\eeann
in terms of their matrix elements reduced with respect to $S,L$ and $T$. The
first four matrix elements, together with the fact that $\langle\bi
w^{(11)}_{d'd'}\rangle =0$ in $d^3$, give equations~\eref{eq:spin1a}
and~\eref{eq:spin1b}.  The last one gives the remaining
relation~\eref{eq:spin1c}.

\section{Concluding remarks}

We started with this relatively simple configuration $d^2d'$ as the subject
of the present work.  The ideas and techniques introduced in this paper can
certainly be applied to other configurations like $d^Nd'$ and $d^Nd'^{N'}$.
Of course, it is a different question whether
these configurations are experimentally observed, or simply of theoretical
interest.  Another obvious extension is to apply it to the $f$-shell
electrons.  The $f$-shell is known to have a richer group structure.
We are aware that similar analysis on the $f$-shell might reveal more
surprises calling for explanations.

Another direction for further research is to introduce the idea of
{\em quasispin} into the analysis.  Quasispin generators are of the type
$(\bi{d^\dag d^\dag})^{(00)}$ and $(\bi{d d})^{(00)}$ in the case of
ordinary $d$-electrons.  These operators, together with
$(\bi{d^\dag d})^{(00)} + (\bi{d d^\dag})^{(00)}$, generate the $SO_Q(3)$
quasispin group.  But in the case where the electrons have an extra spin
structure, namely the isospin, it is known that the analogous quasispin
group is qualitatively different (see Flowers and Szpikowski 1964, Feng and
Judd 1982). The full quasispin group is $SO(8)$ rather than merely $SO(3)$.
This extra group structure may gives us some new selection
rules and proportionality relations, as we know it does in the case of
configurations of equivalent electrons.

\ack
It is a pleasure to thank Prof B R Judd for initiating the work and giving
invaluable advice.  Useful
discussions with Dr M Godefroid have also been much appreciated.

\References
\item[] Bauche J and Klapisch M 1972 \JPB {\bf5} 29-36
\item[] Condon E U and Odabasi H 1980 {\it Atomic Structure}
(Cambridge: Cambridge University Press) pp~583-4
\item[] Condon E U and Shortley G H 1953 {\it The Theory of Atomic Spectra}
(Cambridge: Cambridge University Press)
\item[] Feng C T and Judd B R 1982 \JPA {\bf 15} 2273-84
\item[] Flowers B H and Szpikowski S 1964
{\it Proc.\ Phys.\ Soc.\ }{\bf 84} 673-9
\item[] Godefroid M, Lievin J and Metz Y 1987 \JPB {\bf20} 3283-96
\item[] Hansen J E, Judd B R, Raassen A J J and Uylings P H M 1997 \PRL
{\bf78} 3078-81
\item[] Judd B R 1967 {\it Second Quantization and Atomic Spectroscopy}
(Baltimore: Johns Hopkins Press) p~33
\item[]\dash 1968 {\it Group Theory in Atomic Spectroscopy} in {\it Group
Theory and its Applications} ed E M Loebl (New York: Academic Press)
\item[]\dash 1997 Private communication
\item[]\dash 1998 {\it Operator Techniques in Atomic Spectroscopy}
(Princeton: Princeton University Press)
\item[] Judd B R, Hansen J E and Raassen A J J 1982 \JPB {\bf15} 1457-72
\item[] Judd B R and Leavitt R C 1986 \JPB {\bf19} 485-99
\item[] Kaniauskas J M, \v Simonis V \v C and Rudzikas Z B 1987 \JPB {\bf 20}
3267-81
\item[] Lindgren I and Morrison J 1981 {\it Atomic Many-Body Theory}
(New York: Springer-Verlag) \S11.2
\item[] Nielson C W and Koster G F 1963 {\it Spectroscopic Coefficients for
the $p^n, d^n$ and $f^n$ configurations} (Cambridge, MA: MIT Press)
\item[] Racah G 1954 {\it Bulletin of the Research Council of Israel} {\bf3}
290-8
\item[] \v Simonis V \v C, Kaniauskas J M and Rudzikas Z B 1984
{\it Int.\,J.\,Quantum Chem.} {\bf 25} 57-62
\item[] Sugar J and Corliss C 1985 Atomic Energy Levels of the Iron-Period
Elements: potassium through nickel {\it J.\,Phys.\,Chem.\,Ref.\,Data} {\bf14}
\item[] Weinberg S 1995 {\it The Quantum Theory of Fields} vol~I
(New York: Cambridge University Press) p~200
\item[] Wigner E P 1937 \PR {\bf51} 106-19
\item[] Wybourne B G 1970 {\it Symmetry Principles and Atomic Spectroscopy}
(New York: Wiley-Interscience) table C-3
\endrefs

\end{document}

%% file: newcom.tex
\newcommand{\be}{\begin{equation}}
\newcommand{\ee}{\end{equation}}
\newcommand{\beann}{\begin{eqnarray*}}
\newcommand{\eeann}{\end{eqnarray*}}
\newcommand{\bea}{\begin{eqnarray}}
\newcommand{\eea}{\end{eqnarray}}
\newcommand{\eb}[1]{\mbox{$\overline e_{#1}$}}
\newcommand{\fb}[1]{\mbox{$\overline f_{#1}$}}
\newcommand{\gb}[1]{\mbox{$\overline g_{#1}$}}

\newcommand{\eh}[1]{\mbox{$\widehat e_{#1}$}}
\newcommand{\Eb}[1]{\mbox{$\overline E^{#1}$}}
\newcommand{\Fb}[1]{\mbox{$\overline F^{#1}$}}
\newcommand{\Gb}[1]{\mbox{$\overline G^{#1}$}}
\newcommand{\et}[1]{\mbox{$\widetilde e_{#1}$}}
\newcommand{\Et}[1]{\mbox{$\widetilde E^{#1}$}}

\newcommand{\urs}[2]{\mbox{$[{#1}]_{#2}$}}
\newcommand{\ursb}[2]{\mbox{$[\overline {#1}]_{#2}$}}
\newcommand{\uro}[2]{\mbox{$[{#1}- \overline{#1}]_{#2}$}}
\newcommand{\uroh}[3]{\mbox{$[{#1}- \overline{#2}]_{#3}$}}
\newcommand{\surs}[3]{\mbox{$^{#1}[{#2}]_{#3}$}}

\newcommand{\suro}[3]{\mbox{$^{#1}[{#2}- \overline{#2}]_{#3}$}}
\newcommand{\suroh}[4]{\mbox{$^{#1}[{#2}- \overline{#3}]_{#4}$}}
\newcommand{\A}[3]{\mbox{$(\widehat {\cal A}_{#3})^{(#1)}_{\ {#2}}$}}

\newcommand{\tj}[6]{\mbox{\arraycolsep=0.5mm $\left(\begin{array}{ccc}
 #1 & #2 & #3 \\ #4 & #5 & #6 \end{array} \right)$ }}

%% file: states.tex
%******************************************
% LaTeX: d2d' coulomb  Spring '98
%****************************************************************

\arraycolsep=0.08cm

%\input newcommand.tex
%\begin{document}

\begin{table}
\caption{Transformation between the new and old basis.\label{tb:states}}
\begin{indented}
\item[]\begin{array}[t]{r c@{=} l l l l l} \br
\multicolumn1{c}{$New basis$} & \multicolumn6{c}{$Old basis in terms of $
|d^2 (^{2S'+1}L'),d'; \, ^{2S+1}L\rangle} \\ \mr
|^2[21][3](30)^2S \rangle & & |(^1D) ^2S \rangle \\
|^4[111][21](21)^2P \rangle & & \frac1{\surd2}|(^1D) ^2P \rangle &
+\surd{\frac7{30}}|(^3P)^2P\rangle &-\frac2{\surd{15}}|(^3F)^2P\rangle && \\
|^2[21][111](11)^2P \rangle & & & \ \surd{\frac8{15}} |(^3P) ^2P \rangle &
+\surd{\frac7{15}} |(^3F) ^2P \rangle & & \\
|^2[21][21](21)^2P\rangle & & \frac1{\surd2}|(^1D) ^2P \rangle &
-\surd{\frac7{30}}|(^3P)^2P\rangle &+\frac2{\surd{15}}|(^3F)^2P\rangle && \\
|^4[111][111](11)^4P \rangle & & -\surd{\frac8{15}}|(^3P) ^4P \rangle &
-\surd{\frac7{15}} |(^3F) ^4P \rangle & & & \\
|^2[21][21](21)^4P  \rangle & & \ \surd{\frac7{15}}|(^3P) ^4P \rangle &
-\surd{\frac8{15}} |(^3F) ^4P \rangle & & & \\
|^4[111][21](10)^2D\rangle &&\frac2{\surd{15}}|(^1S)^2D\rangle &-\frac1
{2\surd3}|(^1D)^2D\rangle &-\frac{\surd3}{2\surd5}|(^1G)^2D\rangle &-\frac
{\surd3}{2\surd5}|(^3P)^2D\rangle &-\frac{\surd7}{2\surd5}|(^3F)^2D\rangle \\
|^4[111][21](21)^2D\rangle &&& \ \frac3{2\surd7}|(^1D)^2D\rangle &-\frac
{\surd5}{2\surd7}|(^1G)^2D\rangle &-\frac{\surd7}{2\surd5}|(^3P)^2D\rangle &
+\frac{\surd3}{2\surd5} |(^3F) ^2D \rangle \\
|^2[21][21](21)^2D\rangle &&& \ \frac3{2\surd7}|(^1D)^2D\rangle &-\frac
{\surd5}{2\surd7}|(^1G)^2D\rangle &+\frac{\surd7}{2\surd5}|(^3P)^2D\rangle &
-\frac{\surd3}{2\surd5} |(^3F) ^2D \rangle \\
|^2[21][21](10)^2D\rangle &&\frac2{\surd{15}}|(^1S)^2D\rangle &-\frac1
{2\surd3}|(^1D)^2D\rangle &-\frac{\surd3}{2\surd5}|(^1G)^2D\rangle &+\frac
{\surd3}{2\surd5}|(^3P)^2D\rangle &+\frac{\surd7}{2\surd5}|(^3F)^2D\rangle \\
|^2[21][3](10)^2D\rangle &&\surd{\frac7{15}}|(^1S)^2D\rangle &+\frac2
{\surd{21}}|(^1D)^2D\rangle &+\frac{2\surd3}{\surd35}|(^1G)^2D\rangle && \\
|^2[21][21](10)^4D  \rangle & & \surd{\frac3{10}}|(^3P) ^4D \rangle &
+\surd{\frac7{10}} |(^3F) ^4D \rangle & & & \\
|^2[21][21](21)^4D  \rangle & & \surd{\frac7{10}}|(^3P) ^4D \rangle &
-\surd{\frac3{10}} |(^3F) ^4D \rangle & & & \\
|^4[111][21](21)^2F \rangle & & -\frac1{\surd7}|(^1D) ^2F \rangle &
-\surd\frac5{14}|(^1G) ^2F \rangle & +\surd{\frac25} |(^3P) ^2F \rangle &
+\frac1{\surd{10}} |(^3F) ^2F \rangle & \\
|^2[21][111](11)^2F \rangle & & & & \ \frac1{\surd5} |(^3P) ^2F \rangle &
-\frac2{\surd5} |(^3F) ^2F \rangle & \\
|^2[21][21](21)^2F  \rangle & & -\frac1{\surd7}|(^1D) ^2F \rangle &
-\surd\frac5{14}|(^1G) ^2F \rangle & -\surd{\frac25} |(^3P) ^2F \rangle &
-\frac1{\surd{10}} |(^3F) ^2F \rangle & \\
|^2[21][3](30)^2F   \rangle & & \ \surd\frac57|(^1D) ^2F \rangle &
-\surd\frac27|(^1G) ^2F \rangle & & & \\
|^4[111][111](11)^4F\rangle & & -\frac1{\surd5}|(^3P) ^4F \rangle &
+\frac2{\surd5} |(^3F) ^4F \rangle & & & \\
|^2[21][21](21)^4F  \rangle & & \ \frac2{\surd5}|(^3P) ^4F \rangle &
+\frac1{\surd5} |(^3F) ^4F \rangle & & & \\
|^4[111][21](21)^2G\rangle &&-\surd\frac5{21}|(^1D)^2G\rangle &+\surd{\frac
{11}{42}}|(^1G)^2G\rangle &+\frac1{\surd2}|(^3F)^2G \rangle && \\
|^2[21][21](21)^2G\rangle &&-\surd\frac5{21}|(^1D)^2G\rangle &+\surd{\frac
{11}{42}}|(^1G)^2G\rangle &-\frac1{\surd2}|(^3F)^2G\rangle && \\
|^2[21][3](30)^2G   \rangle & & \ \surd{\frac{11}{21}}|(^1D) ^2G \rangle &
+\surd{\frac{10}{21}}|(^1G) ^2G \rangle & & & \\
|^2[21][21](21)^4G  \rangle & & |(^3F) ^4G\rangle & & & & \\
|^4[111][21](21)^2H\rangle &&\frac1{\surd2}|(^1G)^2H\rangle &
-\frac1{\surd2} |(^3F) ^2H\rangle & & & \\
|^2[21][21](21)^2H\rangle & & \frac1{\surd2}|(^1G) ^2H\rangle &
+\frac1{\surd2} |(^3F) ^2H\rangle & & & \\
|^2[21][21](21)^4H  \rangle & & |(^3F) ^4H\rangle & & & & \\
|^2[21][3](30)^2I   \rangle & & |(^1G) ^2I\rangle & & & & \vspace{3ex} \\
|^2[21]\langle21\rangle(10)^2D\rangle &&& \ \frac52\surd\frac3{77}|(^1D)^2D
\rangle &+\frac32\surd\frac{15}{77}|(^1G) ^2D \rangle & -\frac{\surd{21}}
{2\surd{55}}|(^3P)^2D\rangle & -\frac7{2\surd{55}} |(^3F) ^2D \rangle \\
|^2[21]\langle1\rangle(10)^2D\rangle && \ \surd\frac{11}{15}|(^1S)^2D\rangle
&+\frac1{\surd{33}}|(^1D)^2D\rangle &+\surd{\frac3{55}}|(^1G)^2D\rangle &
+\surd{\frac3{55}}|(^3P)^2D\rangle &+\surd{\frac7{55}}|(^3F)^2D \rangle \\
\br
\end{array}

\item[]\noindent For states belonging to \urs{111}{10}, we use the same
phases as in $d^3$; otherwise, we make an arbitrary phase choice when there
is no precedent to guide us.

\end{indented}
\end{table}

%\end{document}

%% file: tbmatrix.tex
%****************************************************************
% LaTeX: d2d' coulomb  Spring '98
%****************************************************************

\arraycolsep= 0.05cm

%\input newcommand.tex
%\begin{document}

\begin{table}
\caption{Matrix elements in the new basis.\label{tb:matrix}}
\begin{indented}
\item[]\begin{array}[t]{r c c c c c c} \br
 & e_0 & \fb0 & \gb0 & e_1 & \fb1 & \gb1 \\ \mr

^2[21][3](30)^2S \, & 1 & 2 & -1 & 2 & 4 & -2 \\

\begin{array}{r} ^4[111][21](21)^2P \\ ^2[21][111](11)^2P \\ ^2[21][21](21)^2P \end{array} &
\begin{array}{c} 1 \\ 1 \\ 1 \end{array} &
\begin{array}{c} 2 \\ 2 \\ 2 \end{array} &
\begin{array}{c} 2 \\ -1 \\ -1 \end{array} &
\left[ \begin{array}{c c c} 1 & 0 & 1 \\
 0 & 0 & 0 \\ 1 & 0 & 1 \end{array} \right] &
\left[ \begin{array}{c c c} 2 & 0 & -1 \\
 0 & 0 & 0 \\ -1 & 0 & 2 \end{array} \right] &
\left[ \begin{array}{c c c} 2 & 0 & -1 \\
 0 & 0 & 0 \\ -1 & 0 & -1 \end{array} \right] \\

\begin{array}{r} ^4[111][111](11)^4P \\ ^2[21][21](21)^4P \end{array} &
\begin{array}{c} 1 \\ 1 \end{array} &
\begin{array}{c} 2 \\ 2 \end{array} &
\begin{array}{c} 2 \\ -1 \end{array} &
\left[ \begin{array}{c c}  0 & 0 \\  0 & 0 \end{array} \right] &
\left[ \begin{array}{c c}  0 & 0 \\  0 & 3 \end{array} \right] &
\left[ \begin{array}{c c}  0 & 0 \\  0 & -3 \end{array} \right] \\

\begin{array}{r} ^4[111][21](10)^2D \\ ^4[111][21](21)^2D \\ ^2[21][21](21)^2D \\
 ^2[21][21](10)^2D \\ ^2[21][3](10)^2D \end{array} &
\begin{array}{c} 1 \\ 1 \\ 1 \\ 1 \\ 1 \end{array} &
\begin{array}{c} 2 \\ 2 \\ 2 \\ 2 \\ 2 \end{array} &
\begin{array}{c} 2 \\ 2 \\ -1 \\ -1 \\ -1 \end{array} &
\left[ \begin{array}{c c c c c} \frac{7}3 & 0 & 0 & \frac{7}3 &
\frac{2\surd7}3 \\ 0 & 1 & 1 & 0 & 0 \\ 0 & 1 & 1 & 0 & 0 \\
\frac{7}3 & 0 & 0 & \frac{7}3 & \frac{2\surd7}3 \\
\frac{2\surd7}3 & 0 & 0 & \frac{2\surd7}3 & \frac{13}3 \end{array} \right] &
\left[ \begin{array}{c c c c c} \frac{14}3 & 0 & 0 & -\frac{7}3 &
-\frac{2\surd7}3 \\ 0 & 2 & -1 & 0 & 0 \\ 0 & -1 & 2 & 0 & 0 \\
-\frac{7}3 & 0 & 0 & \frac{14}3 & -\frac{2\surd7}3 \\
-\frac{2\surd7}3 & 0 & 0 & -\frac{2\surd7}3 & \frac{26}3 \end{array} \right] &
\left[ \begin{array}{c c c c c} \frac{14}3 & 0 & 0 & -\frac{7}3 &
-\frac{2\surd7}3 \\ 0 & 2 & -1 & 0 & 0 \\ 0 & -1 & -1 & 0 & 0 \\
-\frac{7}3 & 0 & 0 & -\frac{7}3 & \frac{4\surd7}3 \\
-\frac{2\surd7}3 & 0 & 0 & \frac{4\surd7}3 & -\frac{13}3 \end{array} \right] \\

\begin{array}{r}^2[21][21](10)^4D \\ ^2[21][21](21)^4D \end{array} &
\begin{array}{c} 1 \\ 1 \end{array} &
\begin{array}{c} 2 \\ 2 \end{array} &
\begin{array}{c} -1 \\ -1 \end{array} &
\left[ \begin{array}{c c}  0 & 0 \\  0 & 0 \end{array} \right] &
\left[ \begin{array}{c c}  7 & 0 \\  0 & 3 \end{array} \right] &
\left[ \begin{array}{c c}  -7 & 0 \\  0 & -3 \end{array} \right] \\

\begin{array}{r} ^4[111][21](21)^2F \\ ^2[21][111](11)^2F \\ ^2[21][21](21)^2F \\ ^2[21][3](30)^2F \end{array} &
\begin{array}{c} 1 \\ 1 \\ 1 \\ 1 \end{array} &
\begin{array}{c} 2 \\ 2 \\ 2 \\ 2 \end{array} &
\begin{array}{c} 2 \\ -1 \\ -1 \\ -1 \end{array} &
\left[ \begin{array}{c c c c} 1 & 0 & 1 & 0 \\ 0 & 0 & 0 & 0 \\
1 & 0 & 1 & 0 \\ 0 & 0 & 0 & 2 \end{array} \right] &
\left[ \begin{array}{c c c c} 2 & 0 & -1 & 0 \\ 0 & 0 & 0 & 0 \\
-1 & 0 & 2 & 0 \\ 0 & 0 & 0 & 4 \end{array} \right] &
\left[ \begin{array}{c c c c} 2 & 0 & -1 & 0 \\ 0 & 0 & 0 & 0 \\
-1 & 0 & -1 & 0 \\ 0 & 0 & 0 & -2 \end{array} \right] \\

\begin{array}{r}^4[111][111](11)^4F \\ ^2[21][21](21)^4F \end{array} &
\begin{array}{c} 1 \\ 1 \end{array} &
\begin{array}{c} 2 \\ 2 \end{array} &
\begin{array}{c} 2 \\ -1 \end{array} &
\left[ \begin{array}{c c}  0 & 0 \\  0 & 0 \end{array} \right] &
\left[ \begin{array}{c c}  0 & 0 \\  0 & 3 \end{array} \right] &
\left[ \begin{array}{c c}  0 & 0 \\  0 & -3 \end{array} \right] \\

\begin{array}{r}^4[111][21](21)^2G \\ ^2[21][21](21)^2G \\ ^2[21][3](30)^2G \end{array} &
\begin{array}{c} 1 \\ 1 \\ 1 \end{array} &
\begin{array}{c} 2 \\ 2 \\ 2 \end{array} &
\begin{array}{c} 2 \\ -1 \\ -1 \end{array} &
\left[ \begin{array}{ccc} 1 & 1 & 0 \\ 1 & 1 & 0 \\ 0 & 0 & 2 \end{array} \right]&
\left[ \begin{array}{ccc} 2 & -1 & 0 \\ -1 & 2 & 0 \\ 0 & 0 & 4 \end{array} \right] &
\left[ \begin{array}{ccc} 2 & -1 & 0 \\ -1 & -1 & 0 \\ 0 & 0 & -2 \end{array} \right] \\

^2[21][21](21)^4G \, & 1 & 2 & -1 & 0 & 3 & -3 \\

\begin{array}{r}^4[111][21](21)^2H \\ ^2[21][21](21)^2H \end{array} &
\begin{array}{c} 1 \\ 1 \end{array} &
\begin{array}{c} 2 \\ 2 \end{array} &
\begin{array}{c} 2 \\ -1 \end{array} &
\left[ \begin{array}{c c}  1 & 1 \\ 1 & 1 \end{array} \right] &
\left[ \begin{array}{c c}  2 & -1 \\ -1 & 2 \end{array} \right] &
\left[ \begin{array}{c c}  2 & -1 \\ -1 & -1 \end{array} \right] \\

^2[21][21](21)^4H \, & 1 & 2 & -1 & 0 & 3 & -3 \\

^2[21][3](30)^2I \, & 1 & 2 & -1 & 2 & 4 & -2 \\ \br

\end{array}
\noindent
The matrix elements $g_0$ from Condon and Odabasi's table possess a variety
of values. In the new basis, the corresponding \gb0 matrix elements are
diagonal and only take the values 2 and $-1$.
\end{indented}
\end{table}

\addtocounter{table}{-1}
\begin{table}
\caption{laterally continued.}
\begin{indented}
\item[]
\begin{array}[t]{c c c} \br
e_2 & \fb2 & \gb2 \\ \mr

-9 & -18 & 9 \\

\left[ \begin{array}{c c c} -2 & 4\surd7 & -7 \\
4\surd7 & 7 & -4\surd7 \\ -7 & -4\surd7 & -2 \end{array} \right] &
\left[ \begin{array}{c c c} -4 & -4\surd7 & 7 \\
-4\surd7 & 14 & 4\surd7 \\ 7 & 4\surd7 & -4 \end{array} \right] &
\left[ \begin{array}{c c c} -4 & -4\surd7 & 7 \\
-4\surd7 & -7 & -8\surd7 \\ 7 & -8\surd7 & 2 \end{array} \right] \\

\left[ \begin{array}{c c} 7 & -4\surd{14} \\ -4\surd{14} & 5 \end{array} \right] &
\left[ \begin{array}{c c} 14 & 4\surd{14} \\ 4\surd{14} & -11 \end{array} \right] &
\left[ \begin{array}{c c} 14 & 4\surd{14} \\ 4\surd{14} & 16 \end{array} \right] \\

\left[ \begin{array}{c c c c c} 0 & 2\surd{21} & -\surd{21} & 0 & 0 \\
2\surd{21} & 4 & -8 & -\surd{21} & -2\surd3 \\
-\surd{21} & -8 & 4 & 2\surd{21} & -2\surd3 \\
0 & -\surd{21} & 2\surd{21} & 0 & 0 \\
0 & -2\surd3 & -2\surd3 & 0 & 0 \end{array} \right] &
\left[ \begin{array}{c c c c c} 0 & 4\surd{21} & \surd{21} & 0 & 0 \\
4\surd{21} & 8 & 8 & \surd{21} & 2\surd3 \\
\surd{21} & 8 & 8 & 4\surd{21} & 2\surd3 \\
0 & \surd{21} & 4\surd{21} & 0 & 0 \\
0 & 2\surd3 & 2\surd3 & 0 & 0 \end{array} \right] &
\left[ \begin{array}{c c c c c} 0 & 4\surd{21} & \surd{21} & 0 & 0 \\
4\surd{21} & 8 & 8 & \surd{21} & 2\surd3 \\
\surd{21} & 8 & -4 & -2\surd{21} & -4\surd3 \\
0 & \surd{21} & -2\surd{21} & 0 & 0 \\
0 & 2\surd3 & -4\surd3 & 0 & 0 \end{array} \right] \\

\left[ \begin{array}{c c} 0 & 3\surd{21} \\ 3\surd{21} & 12 \end{array} \right] &
\left[ \begin{array}{c c} 0 & 3\surd{21} \\ 3\surd{21} & 0 \end{array} \right] &
\left[ \begin{array}{c c} 0 & 0 \\ 0 & 12 \end{array} \right] \\

\left[ \begin{array}{c c c c} 8 & 6\surd2 & -7 & 2\surd5 \\
6\surd2 & -3 & -6\surd2 & 0 \\ -7 & -6\surd2 & 8 & 2\surd5 \\
2\surd5 & 0 & 2\surd5 & -5 \end{array} \right] &
\left[ \begin{array}{c c c c} 16 & -6\surd2 & 7 & -2\surd5 \\
-6\surd2 & -6 & 6\surd2 & 0 \\ 7 & 6\surd2 & 16 & -2\surd5  \\
-2\surd5 & 0 & -2\surd5 & -10 \end{array} \right] &
\left[ \begin{array}{c c c c} 16 & -6\surd2 & 7 & -2\surd5 \\
-6\surd2 & 3 & -12\surd2 & 0 \\ 7 & -12\surd2 & -8 & 4\surd5 \\
-2\surd5 & 0 & 4\surd5 & 5 \end{array} \right] \\

\left[ \begin{array}{c c} -3 & -12 \\ -12 & 15 \end{array} \right] &
\left[ \begin{array}{c c} -6 & 12 \\ 12 & 9 \end{array} \right] &
\left[ \begin{array}{c c} -6 & 12 \\ 12 & 6 \end{array} \right] \\

\frac13 \ \left[ \begin{array}{c c c} -16 & 11 & 2\surd{55} \\
11 & -16 & 2\surd{55} \\ 2\surd{55} & 2\surd{55} & -7 \end{array} \right] &
\frac13 \ \left[ \begin{array}{c c c} -32 & -11 & -2\surd{55} \\
-11 & -32 & -2\surd{55} \\ -2\surd{55} & -2\surd{55} & -14 \end{array} \right] &
\frac13 \ \left[ \begin{array}{c c c} -32 & -11 & -2\surd{55} \\
-11 & 16 & 4\surd{55} \\ -2\surd{55} & 4\surd{55} & 7 \end{array} \right] \\

-9 & -7 & -2 \\

\left[ \begin{array}{c c} -2 & 7 \\ 7 & -2 \end{array} \right] &
\left[ \begin{array}{c c} -4 & -7 \\ -7 & -4 \end{array} \right] &
\left[ \begin{array}{c c} -4 & -7 \\ -7 & 2 \end{array} \right] \\

-9 & 3 & -12 \\

5 & 10 & -5 \\ \br

\end{array}
\end{indented}
\end{table}

%\end{document}

%% file: ubranch.tex
%****************************************************************
% LaTeX: d2d' coulomb  Spring '98
%****************************************************************

\arraycolsep=.15cm

%\input newcommand.tex
%\begin{document}

\begin{table}
\caption{Branching rule $U(2n) \rightarrow SO(3) \times U(n)$.\label{tb:ugpbr}}
\begin{indented}
\item[]\begin{array}[t]{l l} \br
U(2n) & SO(3) \times U(n) \\ \mr
\urs1{2n} & \surs21n \\
\urs{11}{2n} & \surs3{11}n + {}\surs12n \\
\hspace{1ex} \vdots & \hspace{3ex} \vdots \\
\urs{1^r}{2n} & \surs{r+1}{1^r}n + {}\surs{r-1}{21^{(r-2)}}n + \cdots \\
\urs2{2n} & \surs1{11}n + {}\surs32n \\
\urs{21}{2n} & \surs2{111}n + {}\surs{2,4}{21}n + {}\surs23n\\
\uro1{2n} & \surs30n + {}\suro{1,3}1n \\
\uro{11}{2n} & \surs{1,5}0n + {}\suro{1,3,3,5}1n + {}\suro{1,3,5}{11}n + {}
\suroh3{11}2n + {}\suroh32{11}n + {}\suro12n \\
\uro2{2n} & \surs{1,5}0n+{}\suro{1,3,3,5}1n+{}\suro1{11}n+{}\suroh3{11}2n+{}
\suroh32{11}n + {}\suro{1,3,5}2n \\
\uroh2{11}{2n} & \surs30n+{}\suro{1,3,3,5}1n+{}\suro3{11}n+{}\suroh1{11}2n+{}
\suroh{1,3,5}2{11}n + {}\suro32n \\ \br

\end{array}
\end{indented}
\end{table}

%\end{document}

%% file: oplabel.tex
%****************************************************************
% LaTeX: d2d' coulomb  Spring '98
%****************************************************************

%\input newcommand.tex
%\begin{document}

\begin{table}
\caption{Group labels of the Coulomb interaction operators.\label{tb:oplabel}}
\begin{indented}
\item[]\begin{tabular}{l l l l} \br
operator & $SO_T(3) \times U(10)$ & $U(5)$ & $SO(5)$ \\ \mr
$e_0$, \eb0  & \surs{1,5}0{10} & \urs05 & (00) \\
$e_1$, \eb1  & \surs{1,5}0{10} & \urs05 & (00) \\
       & \suro{1,5}{11}{10} & \urs05 + \uro25 & (00) \\
%$e'_1$, \ebp & \suro{1,5}{11}{10} & \urs05 + \uro25 & (00) \\
$e_2$, \eb2  & \suro{1,5}{11}{10} & \uro{11}5 + \uro25 & (22) \\
\fb0, \gb0 & \surs{1,5}0{10} & \urs05 & (00) \\
\fb1, \gb1 & \surs{1,5}0{10} & \urs05 & (00) \\
           & \suro{1,5}{11}{10} & \urs05 + \uro25 & (00) \\
           & \suro12{10} & \urs05 + \uro25 & (00) \\
%\fbp, \gbp & \suro{1,5}{11}{10} & \urs05 + \uro25 & (00) \\
%           & \suro12{10} & \urs05 + \uro25 & (00) \\
\fb2, \gb2 & \suro{1,5}{11}{10} & \uro{11}5 + \uro25 & (22) \\
           & \suro12{10} & \uro{11}5 + \uro25 & (22) \\
\et0 & \surs10{10} & \urs05 & (00) \\
\et1 & \surs10{10} & \urs05 & (00) \\
     & \suro12{10} & \urs05 + \uro25 & (00) \\
%\etp & \suro12{10} & \urs05 + \uro25 & (00) \\
\et2 & \suro12{10} & \uro{11}5 + \uro25 & (22) \\
$\eb0 + e_0$ & \surs10{10} & \urs05 & (00) \\
$\eb1 + e_1$ & \surs10{10} & \urs05 & (00) \\
              & \suro1{11}{10} & \urs05 + \uro25 & (00) \\
%$\ebp +e'_1$ & \suro1{11}{10} & \urs05 + \uro25 & (00) \\
$\eb2 + e_2$ & \suro1{11}{10} & \uro{11}5 + \uro25 & (22) \\
$2\eb0 - e_0$ & \surs50{10} & \urs05 & (00) \\
$2\eb1 - e_1$ & \surs50{10} & \urs05 & (00) \\
               & \suro5{11}{10} & \urs05 + \uro25 & (00) \\
%$2\ebp -e'_1$ & \suro5{11}{10} & \urs05 + \uro25 & (00) \\
$2\eb2 - e_2$ & \suro5{11}{10} & \uro{11}5 + \uro25 & (22) \\ \br
\end{tabular}
\end{indented}
\end{table}

%\end{document}

%% file: tbnewmat.tex
%****************************************************************
% LaTeX: d2d' coulomb  Spring '98
%****************************************************************

\arraycolsep= 0.05cm

%\input newcommand.tex
%\begin{document}

\begin{table}
\caption{Matrix elements for the operators $e_i$, \eb i, \et i and \eh i.
\label{tb:newmatrix}}
\begin{indented}
\item[]\begin{array}[t]{r c c c c c c c c} \br
 & e_0 & \eb0 & \et0 & \eh0 & e_1 & \eb1 & \et1 & \eh1 \\ \mr

^2[21][3](30)^2S \, & 1 & \frac12 & \frac32 & 3 & 2 & 1 & 3 & 6 \\

\begin{array}{r} ^4[111][21](21)^2P \\ ^2[21][111](11)^2P \\ ^2[21][21](21)^2P \end{array} &
\begin{array}{c} 1 \\ 1 \\ 1 \end{array} &
\begin{array}{c} 2 \\ \frac12 \\ \frac12 \end{array} &
\begin{array}{c} 0 \\ \frac32 \\ \frac32 \end{array} &
\begin{array}{c} 3 \\ 3 \\ 3 \end{array} &
\left[ \begin{array}{c c c} 1 & 0 & 1 \\
 0 & 0 & 0 \\ 1 & 0 & 1 \end{array} \right] &
\left[ \begin{array}{c c c} 2 & 0 & -1 \\
 0 & 0 & 0 \\ -1 & 0 & \frac12 \end{array} \right] &
\left[ \begin{array}{c c c} 0 & 0 & 0 \\
 0 & 0 & 0 \\ 0 & 0 & \frac32 \end{array} \right] &
\begin{array}{c} 3 \\ 0 \\ 3 \end{array} \\

\begin{array}{r}^4[111][111](11)^4P \\ ^2[21][21](21)^4P \end{array} &
\begin{array}{c} 1 \\ 1 \end{array} &
\begin{array}{c} 2 \\ \frac12 \end{array} &
\begin{array}{c} 0 \\ \frac32 \end{array} &
\begin{array}{c} 3 \\ 3 \end{array} &
\left[ \begin{array}{c c}  0 & 0 \\  0 & 0 \end{array} \right] &
\left[ \begin{array}{c c}  0 & 0 \\  0 & 0 \end{array} \right] &
\left[ \begin{array}{c c}  0 & 0 \\  0 & 3 \end{array} \right] &
\begin{array}{c} 0 \\ 3 \end{array} \\

\begin{array}{r}^4[111][21](10)^2D \\ ^4[111][21](21)^2D \\ ^2[21][21](21)^2D \\
 ^2[21][21](10)^2D \\ ^2[21][3](10)^2D \end{array} &
\begin{array}{c} 1 \\ 1 \\ 1 \\ 1 \\ 1 \end{array} &
\begin{array}{c} 2 \\ 2 \\ \frac12 \\ \frac12 \\ \frac12 \end{array} &
\begin{array}{c} 0 \\ 0 \\ \frac32 \\ \frac32 \\ \frac32 \end{array} &
\begin{array}{c} 3 \\ 3 \\ 3 \\ 3 \\ 3 \end{array} &
\left[ \begin{array}{c c c c c} \frac73 & 0 & 0 & \frac73 &
\frac{2\surd7}3 \\ 0 & 1 & 1 & 0 & 0 \\ 0 & 1 & 1 & 0 & 0 \\
\frac73 & 0 & 0 & \frac73 & \frac{2\surd7}3 \\
\frac{2\surd7}3 & 0 & 0 & \frac{2\surd7}3 & \frac{13}3 \end{array} \right] &
\left[ \begin{array}{c c c c c} \frac{14}3 & 0 & 0 & -\frac73 &
-\frac{2\surd7}3 \\ 0 & 2 & -1 & 0 & 0 \\ 0 & -1 & \frac12 & 0 & 0 \\
-\frac73 & 0 & 0 & \frac76 & \frac{\surd7}3 \\
-\frac{2\surd7}3& 0 & 0 & \frac{\surd7}3 & \frac{13}6 \end{array} \right] &
\left[ \begin{array}{c c c c c} 0 & 0 & 0 & 0 &
0 \\ 0 & 0 & 0 & 0 & 0 \\ 0 & 0 & \frac32 & 0 & 0 \\
0 & 0 & 0 & \frac72 & -\surd7 \\
0 & 0 & 0 & -\surd7 & \frac{13}2 \end{array} \right] &
\begin{array}{c} 7 \\ 3 \\ 3 \\ 7 \\ 13 \end{array} \\

\begin{array}{r}^2[21][21](10)^4D \\ ^2[21][21](21)^4D \end{array} &
\begin{array}{c} 1 \\ 1 \end{array} &
\begin{array}{c} \frac12 \\ \frac12 \end{array} &
\begin{array}{c} \frac32 \\ \frac32 \end{array} &
\begin{array}{c} 3 \\ 3 \end{array} &
\left[ \begin{array}{c c}  0 & 0 \\  0 & 0 \end{array} \right] &
\left[ \begin{array}{c c}  0 & 0 \\  0 & 0 \end{array} \right] &
\left[ \begin{array}{c c}  7 & 0 \\  0 & 3 \end{array} \right] &
\begin{array}{c} 7 \\ 3 \end{array} \\

\begin{array}{r} ^4[111][21](21)^2F \\ ^2[21][111](11)^2F \\ ^2[21][21](21)^2F \\ ^2[21][3](30)^2F \end{array} &
\begin{array}{c} 1 \\ 1 \\ 1 \\ 1 \end{array} &
\begin{array}{c} 2 \\ \frac12 \\ \frac12 \\ \frac12 \end{array} &
\begin{array}{c} 0 \\ \frac32 \\ \frac32 \\ \frac32 \end{array} &
\begin{array}{c} 3 \\ 3 \\ 3 \\ 3 \end{array} &
\left[ \begin{array}{c c c c} 1 & 0 & 1 & 0 \\ 0 & 0 & 0 & 0 \\
1 & 0 & 1 & 0 \\ 0 & 0 & 0 & 2 \end{array} \right] &
\left[ \begin{array}{c c c c} 2 & 0 & -1 & 0 \\ 0 & 0 & 0 & 0 \\
-1 & 0 & \frac12 & 0 \\ 0 & 0 & 0 & 1 \end{array} \right] &
\left[ \begin{array}{c c c c} 0 & 0 & 0 & 0 \\ 0 & 0 & 0 & 0 \\
0 & 0 & \frac32 & 0 \\ 0 & 0 & 0 & 3 \end{array} \right] &
\begin{array}{c} 3 \\ 0 \\ 3 \\ 6 \end{array} \\

\begin{array}{r}^4[111][111](11)^4F \\ ^2[21][21](21)^4F \end{array} &
\begin{array}{c} 1 \\ 1 \end{array} &
\begin{array}{c} 2 \\ \frac12 \end{array} &
\begin{array}{c} 0 \\ \frac32 \end{array} &
\begin{array}{c} 3 \\ 3 \end{array} &
\left[ \begin{array}{c c}  0 & 0 \\  0 & 0 \end{array} \right] &
\left[ \begin{array}{c c}  0 & 0 \\  0 & 0 \end{array} \right] &
\left[ \begin{array}{c c}  0 & 0 \\  0 & 3 \end{array} \right] &
\begin{array}{c} 0 \\ 3 \end{array} \\

\begin{array}{r}^4[111][21](21)^2G \\ ^2[21][21](21)^2G \\ ^2[21][3](30)^2G \end{array} &
\begin{array}{c} 1 \\ 1 \\ 1 \end{array} &
\begin{array}{c} 2 \\ \frac12 \\ \frac12 \end{array} &
\begin{array}{c} 0 \\ \frac32 \\ \frac32 \end{array} &
\begin{array}{c} 3 \\ 3 \\ 3 \end{array} &
\left[ \begin{array}{ccc} 1 & 1 & 0 \\ 1 & 1 & 0 \\ 0 & 0 & 2 \end{array} \right] &
\left[ \begin{array}{ccc} 2 & -1 & 0 \\ -1 & \frac12 & 0 \\ 0 & 0 & 1 \end{array} \right] &
\left[ \begin{array}{ccc} 0 & 0 & 0 \\ 0 & \frac32 & 0 \\ 0 & 0 & 3 \end{array} \right] &
\begin{array}{c} 3 \\ 3 \\ 6 \end{array} \\

^2[21][21](21)^4G \, & 1 & \frac12 & \frac32 & 3 & 0 & 0 & 3 & 3 \\

\begin{array}{r}^4[111][21](21)^2H \\ ^2[21][21](21)^2H \end{array} &
\begin{array}{c} 1 \\ 1 \end{array} &
\begin{array}{c} 2 \\ \frac12 \end{array} &
\begin{array}{c} 0 \\ \frac32 \end{array} &
\begin{array}{c} 3 \\ 3 \end{array} &
\left[ \begin{array}{c c}  1 & 1 \\ 1 & 1 \end{array} \right] &
\left[ \begin{array}{c c}  2 & -1 \\ -1 & \frac12 \end{array} \right] &
\left[ \begin{array}{c c}  0 & 0 \\ 0 & \frac32 \end{array} \right] &
\begin{array}{c} 3 \\ 3 \end{array} \\
		
^2[21][21](21)^4H \, & 1 & \frac12 & \frac32 & 3 & 0 & 0 & 3 & 3 \\

^2[21][3](30)^2I \, & 1 & \frac12 & \frac32 & 3 & 2 & 1 & 3 & 6 \\ \br
\end{array}
\end{indented}
\end{table}

\addtocounter{table}{-1}\begin{table}
\caption{laterally continued.}
\begin{indented}
\item[]\begin{array}[t]{c c c c} \br
e_2 & \eb2 & \et2 & \eh2 \\ \mr

-9 & -\frac92 & -\frac{27}2 & -27 \\

\left[ \begin{array}{c c c} -2 & 4\surd7 & -7 \\
 4\surd7 & 7 & -4\surd7 \\ -7 & -4\surd7 & -2 \end{array} \right] &
\left[ \begin{array}{c c c} -4 & -4\surd7 & 7 \\
-4\surd7 & \frac72 & -2\surd7 \\ 7 & -2\surd7 & -1 \end{array} \right] &
\left[ \begin{array}{c c c} 0 & 0 & 0 \\
0 & \frac{21}2 & 6\surd7 \\ 0 & 6\surd7 & -3 \end{array} \right] &
\begin{array}{c} -6 \\ 21 \\ -6 \end{array} \\

\left[ \begin{array}{c c} 7 & -4\surd{14} \\ -4\surd{14} & 5 \end{array} \right] &
\left[ \begin{array}{c c} 14 & 4\surd{14} \\ 4\surd{14} & \frac52 \end{array} \right] &
\left[ \begin{array}{c c} 0 & 0 \\ 0 & -\frac{27}2 \end{array} \right] &
\begin{array}{c} 21 \\ -6 \end{array} \\

\left[ \begin{array}{c c c c c} 0 & 2\surd{21} & -\surd{21} & 0 & 0 \\
2\surd{21} & 4 & -8 & -\surd{21} & -2\surd3 \\
-\surd{21} & -8 & 4 & 2\surd{21} & -2\surd3 \\
0 & -\surd{21} & 2\surd{21} & 0 & 0 \\
0 & -2\surd3 & -2\surd3 & 0 & 0 \end{array} \right] &
\left[ \begin{array}{c c c c c} 0 & 4\surd{21} & \surd{21} & 0 & 0 \\
4\surd{21} & 8 & 8 & \surd{21} & 2\surd3 \\
\surd{21} & 8 & 2 & \surd{21} & -\surd3 \\
0 & \surd{21} & \surd{21} & 0 & 0 \\
0 & 2\surd3 & -\surd3 & 0 & 0 \end{array} \right] &
\left[ \begin{array}{c c c c c} 0 & 0 & 0 & 0 & 0 \\
0 & 0 & 0 & 0 & 0 \\
0 & 0 & 6 & 3\surd{21} & 3\surd3 \\
0 & 0 & 3\surd{21} & 0 & 0 \\
0 & 0 & 3\surd3 & 0 & 0 \end{array} \right] &
\begin{array}{c}
\left[ \begin{array}{c c} 0 & 6\surd{21} \\ 6\surd{21} & 12 \end{array} \right] \\
\left[ \begin{array}{c c} 12 & 6\surd{21} \\ 6\surd{21} & 0 \end{array} \right] \\
0 \end{array} \\

\left[ \begin{array}{c c} 0 & 3\surd{21} \\ 3\surd{21} & 12 \end{array} \right] &
\left[ \begin{array}{c c} 0 & \frac{3\surd{21}}2 \\ \frac{3\surd{21}}2 & 6 \end{array} \right] &
\left[ \begin{array}{c c} 0 & \frac{3\surd{21}}2 \\ \frac{3\surd{21}}2 & -6 \end{array} \right] &
\left[ \begin{array}{c c} 0 & 6\surd{21} \\ 6\surd{21} & 12 \end{array} \right] \\

\left[ \begin{array}{c c c c} 8 & 6\surd2 & -7 & 2\surd5 \\
6\surd2 & -3 & -6\surd2 & 0 \\ -7 & -6\surd2 & 8 & 2\surd5 \\
2\surd5 & 0 & 2\surd5 & -5 \end{array} \right] &
\left[ \begin{array}{c c c c} 16 & -6\surd2 & 7 & -2\surd5 \\
-6\surd2 & -\frac32 & -3\surd2 & 0 \\ 7 & -3\surd2 & 4 & \surd5 \\
-2\surd5 & 0 & \surd5 & -\frac52 \end{array} \right] &
\left[ \begin{array}{c c c c} 0 & 0 & 0 & 0 \\ 0 & -\frac92 & 9\surd2 & 0 \\
0 & 9\surd2 & 12 & -3\surd5 \\ 0 & 0 & -3\surd5 & -\frac{15}2 \end{array} \right] &
\begin{array}{c} 24 \\ -9 \\ 24 \\ -15 \end{array} \\

\left[ \begin{array}{c c} -3 & -12 \\ -12 & 15 \end{array} \right] &
\left[ \begin{array}{c c} -6 & 12 \\ 12 & \frac{15}2 \end{array} \right] &
\left[ \begin{array}{c c} 0 & 0 \\ 0 & \frac32 \end{array} \right] &
\begin{array}{c} -9 \\ 24 \end{array} \\

\frac13 \ \left[ \begin{array}{c c c} -16 & 11 & 2\surd{55} \\
 11 & -16 & 2\surd{55} \\ 2\surd{55} & 2\surd{55} & -7 \end{array} \right] &
\frac13 \ \left[ \begin{array}{c c c} -32 & -11 & -2\surd{55} \\
-11 & -8 & \surd{55} \\ -2\surd{55} & \surd{55} & -\frac72 \end{array} \right] &
\frac13 \ \left[ \begin{array}{c c c} 0 & 0 & 0 \\
0 & -24 & -3\surd{55} \\ 0 & -3\surd{55} & -\frac{21}2 \end{array} \right] &
\begin{array}{c} -16 \\ -16 \\ -7 \end{array} \\

-9 & -\frac92 & -\frac52 & -16 \\

\left[ \begin{array}{c c} -2 & 7 \\ 7 & -2 \end{array} \right] &
\left[ \begin{array}{c c} -4 & -7 \\ -7 & -1 \end{array} \right] &
\left[ \begin{array}{c c} 0 & 0 \\ 0 & -3 \end{array} \right] &
\begin{array}{c} -6 \\ -6 \end{array} \\

-9 & -\frac92 & \frac{15}2 & -6 \\

5 & \frac52 & \frac{15}2 & 15 \\ \br

\end{array}
\end{indented}
\end{table}

%\end{document}

%% file: ucross.tex
%****************************************************************
% LaTeX: Unitrary Group Kronecker product table  Spring '98
%****************************************************************

\arraycolsep=.15cm

%\input newcommand.tex
%\begin{document}

\begin{table}
\caption{Some Kronecker products for Unitary Groups $U(n)$.\label{tb:ucross}}
\begin{indented}
\item[]\begin{tabular}[t]{l l} \br
\urs1{}$\times$\urs1{} &  \urs2{} + \urs{11}{} \\
\urs1{}$\times$\ursb1{} &  \uro1{} + \urs0{} \\
\urs{11}{}$\times$\urs{11}{} & \urs{22}{} + \urs{211}{} + \urs{1^4}{} \\
\urs{11}{}$\times$\ursb{11}{} & \uro{11}{} + \uro1{} + \urs0{} \\
\urs{11}{}$\times$\urs2{} & \urs{31}{} + \urs{211}{} \\
\urs{11}{}$\times$\ursb2{} & \uroh{11}2{} + \uro1{} \\
\urs2{}$\times$\urs2{} & \urs4{} + \urs{31}{} + \urs{22}{} \\
\urs2{}$\times$\ursb2{} & \uro2{} + \uro1{} + \urs0{} \\
\urs{111}{}$\times$\urs{111}{} & \urs{222}{} + \urs{2211}{} + \urs{21^4}{} + \urs{1^6}{} \\
\urs{111}{}$\times$\ursb{111}{} & \uro{111}{} + \uro{11}{} + \uro1{} + \urs0{} \\
\urs{111}{}$\times$\urs{21}{} & \urs{321}{} + \urs{3111}{} + \urs{2211}{} + \urs{21^4}{} \\
\urs{111}{}$\times$\ursb{21}{} & \uroh{111}{12}{} + \uroh{11}2{} + \uro{11}{} + \uro1{} \\
\urs{111}{}$\times$\urs3{} & \urs{411}{} + \urs{3111}{} \\
\urs{111}{}$\times$\ursb3{} & \uroh{111}3{} + \uroh{11}2{} \\
\urs{21}{}$\times$\urs{21}{} & \urs{42}{} + \urs{411}{} + \urs{33}{} + 2\urs{321}{} + \urs{3111}{} + \urs{222}{} + \urs{2211}{} \\
\urs{21}{}$\times$\ursb{21}{} & \uroh{21}{12}{} + \uroh{11}2{} + \uroh2{11}{} + \uro2{} + \uro{11}{} + 2\uro1{} + \urs0{} \\
\urs{21}{}$\times$\urs3{} & \urs{51}{} + \urs{42}{} + \urs{411}{} + \urs{321}{} \\
\urs{21}{}$\times$\ursb3{} & \uroh{21}3{} + \uroh{11}2{} + \uro2{} + \uro1{} \\
\urs3{}$\times$\urs3{} & \urs6{} + \urs{51}{} + \urs{42}{} + \urs{33}{} \\
\urs3{}$\times$\ursb3{} & \uro3{} + \uro2{} + \uro1{} + \urs0{} \\
\br
\end{tabular}

\item[] The above products are obtained using the method of Young's
tableaux.  When negative integers are involved in the irrep, we first add a
constant to it so that all integers becomes non-negative; then use the
standard Young's method to find the Kronecker product, and finally subtract
the same constant from the resulting irreps.

\item[] The Kronecker products are valid for large enough $n$.
For instance, \uro{111}5 contains at least six entries and is identically
zero for $U(5)$.

\end{indented}
\end{table}

\arraycolsep=.05cm

%\end{document}

%% file: casimir.tex
%****************************************************************
% LaTeX: d2d' coulomb  Spring '98
%****************************************************************

%\input newcommand.tex
%\begin{document}

\begin{table}
\caption{Eigenvalues of the $U(10)$ scalar operators.\label{tb:casimir}}
\begin{indented}
\item[]\begin{tabular}{l l} \br
operator & eigenvalue \\ \mr
$e_0$  & $\frac12 N(N-1) + \frac12 N'(N'-1)$ \\
\fb{0} & $N N'$ \\
\gb{0} & $:T_+T_- \! : \; = T^2 - T_0^2 + T_0 - N$ \\
       &  $= T^2 - (\frac{N+N'}2)(\frac{N+N'}2+1) + NN'$ \\
\eb{0} & $\frac12 \{ T^2 - (\frac{N+N'}2)(\frac{N+N'}2+1) \} + NN' $ \\
\et{0} & $\frac12 \{-T^2 + (\frac{N+N'}2)(\frac{N+N'}2+1) \}$ \\
$\eb0 + e_0$ & $\frac12 \{ T^2 + 3 (\frac{N+N'}2)(\frac{N+N'}2-1) \}$ \\
$2\eb0- e_0$ & $T^2 - 3 T_0^2$ \\ \br
\end{tabular}
\end{indented}
\end{table}

%\end{document}